\documentclass[preprint,groupedaddress,floatfix,%
nofootinbib]{revtex4}

\usepackage{euscript}
\usepackage{dcolumn}% Align table columns on decimal point
\usepackage{graphicx}
\usepackage{epsfig}
\usepackage{hyperref}
\usepackage{graphicx}% Include figure files
\usepackage{dcolumn}% Align table columns on decimal point
\usepackage{longtable}
\usepackage{times}
\usepackage{amsmath,amssymb,bm}
\usepackage[english]{babel}
\usepackage[latin1]{inputenc}
\usepackage{times}
\usepackage[T1]{fontenc}
\usepackage{color}
\usepackage{fancybox}
\usepackage{sidecap}
\usepackage{graphicx}
\usepackage{epsfig}
\usepackage{psfrag}
\usepackage{bbm}
\usepackage[english]{babel}
\usepackage[latin1]{inputenc}
\usepackage{times}
\usepackage[T1]{fontenc}
\usepackage{color}
\usepackage{fancybox}
\usepackage{sidecap}
\usepackage{psfrag}
\usepackage{bbm}
\usepackage{wasysym}
\usepackage{slashed}
\usepackage{tikz}
\usepackage{euscript}

\usepackage{framed}
\begin{document}
%\begin{CJK*}{GBK}{hei}

\title{Deterministic Preparation of Arbitrary Spin Eigenfunctions}
\author{Wenxuan Tao}
\author{Jianan Wang}
\author{Fen Zuo\footnote{Email: \textsf{zuofen@miqroera.com}}}
\affiliation{Shanghai MiQro Era Digital Technology Co. Ltd., Shanghai, China}

\begin{abstract}
Quantum states with conserved total spins, or spin eigenfunctions, are important for studying quantum chemistry and quantum manybody physics
problems. A typical class of spin eigenfunctions are Dicke states, which attain maximal spins.
While we already have many efficient quantum algorithms to prepare Dicke states, it is not yet clear if we could do so for
arbitrary spin eigenfunctions deterministically. Generalizing B\"{a}rtschi and Eidenbenz's elegant algorithms for Dicke state preparation,
we successfully prepare arbitrary spin eigenfunctions characterized by branching paths and binary spin trees.
%The canonical spin eigenfunctions characterized
%by branching diagram symbols fall within this class, and are prepared and studied separately.
As a byproduct, we also develop the corresponding classical algorithms to reconstruct all these spin states.
\end{abstract}

\maketitle

\tableofcontents

\section{Motivation}

A spin-$1/2$ particle is born to be a qubit. When many elementary spins are coupled together, we would like to study those states with
definite total spins. These are called spin eigenfunctions in the quantum chemistry literature. How could we study these states in the
quantum computing framework?

Among these spin eigenfunctions are a special class of states, the so-called Dicke states.
These are simply spin eigenfunctions with the maximal spins. They have received a lot of attention in recent years.
For example, they are used as the initial states for the recently proposed algorithm for optimization,
Decoded Quantum Interferometry~(DQI)\cite{DQI-2024}.
There are many quantum algorithms for preparing Dicke states. \cite{BE-2022} provides a nice summary of various preparation
schemes for Dicke states and the related symmetric states. Among all these, the two algorithms presented in \cite{BE-2019}
and \cite{BE-2022} are very succinct and elegant, and will be discussed in detail later.

An interesting question then arises: could we generalize the preparation circuits of Dicke states to arbitrary spin eigenfunctions?
The question may sound a little naive, if we take Feynman's original proposal for quantum computation seriously.
Indeed, since all spin eigenfunctions appear naturally in realistic physical systems, it would be quite strange if they could not
be efficiently prepared on quantum computers. Nevertheless, a preliminary analysis in~\cite{Carbone2022} concludes that
the required gates for preparing general spin eigenfunctions would be exponentially increasing with the particle number.
If this conclusion is valid universally, then how could all the efficient algorithms for Dicke states come into existence?
If you compare the derivation in~\cite{Carbone2022} with the explicit algorithms in \cite{BE-2019} and \cite{BE-2022},
you will find a crucial difference. While \cite{Carbone2022} uses nested loops to deal with states with different spin projections,
both \cite{BE-2019} and \cite{BE-2022} manage to handle them in parallel. This difference immediately gives rise to the dramatically
different complexities. Then, which one of them is more general? Or, could we generalize the parallelization technique in \cite{BE-2019}
and \cite{BE-2022} to arbitrary spin eigenfunctions?

It is believed~\cite{Marti-2025} that the parallelization techniques in \cite{BE-2019} and \cite{BE-2022} heavily rely on the
symmetry of Dicke states. Indeed, Dicke states are totally symmetric states, and all the components contribute equally.
This simplifies the coupling of Dicke states to combinatoric problems. Actually, the key relations in both \cite{BE-2019}
and \cite{BE-2022} are derived based on this. However, if we translate the whole framework from the coding language to the
spin language, these relations could be easily generalized with the help of the powerful angular momentum theory. Therefore,
it seems that symmetry would never be the obstacle to generalizing those algorithms.

When we have all these in mind, the generalization is in fact quite straightforward.
All we need to do is to translate everything from the coding language to the spin language, and then adapt it to the
general picture. We will illustrate the whole procedure as follows. In the next section, we introduce the two algorithms
for preparing Dicke states in \cite{BE-2019} and \cite{BE-2022}. Then in section III and IV, we show how to adapt them to
general spin eigenfunctions, respectively. In section V, we give the corresponding classical algorithms for verification.
We summarize all the results in the final section, and suggest some future directions.

\section{Dicke states}

In this section we will review the two elegant algorithms for preparing Dicke states proposed by B\"{a}rtschi and Eidenbenz
in \cite{BE-2019} and \cite{BE-2022}. Before doing so, we would like to show explicitly how Dicke states could be identified
with spin eigenfunctions of maximal possible spins. We believe this will help the readers get acquainted with the spin language.

\subsection{Dicke states}

As we said before, a spin-$1/2$ particle is born a qubit. Let us make this relation explicit first. We define the single-particle spin vector as:
\begin{equation}
\vec S:=(\hat S^x,\hat S^y,\hat S^z)=(\frac{X}{2},\frac{Y}{2},\frac{Z}{2}).
\end{equation}
Then $\vec S^2$ and $\hat S^z$ commute, and could be diagonalized simultaneously. Let us denote the eigenvalue of $\vec S^2$ as $S(S+1)$,
and that of $\hat S^z$ as $S^z$. Since
\begin{equation}
\vec S^2:= \vec S \cdot \vec S=\frac{3}{4}=\frac{1}{2}(\frac{1}{2}+1),
\end{equation}
we have $S=1/2$. Thus we get two orthogonal states $|S=1/2, S^z=\pm1/2\rangle$, which could be identified with the single-qubit states as:
\begin{equation}
|S=\frac{1}{2},S^z=\frac{1}{2}\rangle \equiv |0\rangle;\quad  |S=\frac{1}{2},S^z=-\frac{1}{2}\rangle \equiv |1\rangle.
\end{equation}
Now we consider the many-particle case. The total spin vector is naturally defined as
\begin{equation}
\hat S^x:=\frac{1}{2}\sum_i^n~X_i,\quad \hat S^y:=\frac{1}{2}\sum_i^n~Y_i, \quad \hat S^z:=\frac{1}{2}\sum_i^n~Z_i.
\end{equation}
We would like to calculate its square $\vec S^2$. Notice that the SWAP gate, which we denote as $F_{ij}$, has the following expansion:
\begin{equation}
F_{ij}=\frac{1}{2}(I_iI_j+X_iX_j+Y_iY_j+Z_iZ_j).\label{eq:SWAP}
\end{equation}
So we could express the square of the total spin operator as:
\begin{equation}
\vec S^2=(\hat S^x)^2+(\hat S^y)^2+(\hat S^z)^2=\frac{n(4-n)}{4}+\sum_{i<j}~F_{ij}.\label{eq:Spin-swap}
\end{equation}
Since all SWAP gates are unitary, we could attempt to make all of them take maximum eigenvalue 1, and get
\begin{equation}
\vec S^2=\frac{n}{2}(\frac{n}{2}+1).
\end{equation}
So the maximum spin we can get is $S=n/2$. The spin projection $S^z$ is simply related to the Hamming weight as $S^z=\frac{n}{2}-k$.
So when the weight $k$ increases from $0$ to $n$, the spin projection $S^z$ decreases from $n/2$ to $-n/2$.
The corresponding state would be stabilized by all SWAP gates, and thus by the whole permutation group ${\mathcal S}_n$ generated by them.
It could be explicitly produced through group averaging as
\begin{eqnarray}
|S=\frac{n}{2},S^z=\frac{n}{2}-k\rangle&\propto &\sum_{\sigma_n \in {\mathcal S}_n} \sigma_n |0^{n-k}\rangle ~|1^{k}\rangle\nonumber\\
&\propto&\sum_{x\in \{0,1\}^n,{|x|}=k}~|x\rangle.
\end{eqnarray}
Here $|\cdot|$ is the Hamming weight. Upon normalization, we get
\begin{equation}
|S=\frac{n}{2},S^z=\frac{n}{2}-k\rangle=\frac{1}{\sqrt{\tbinom{n}{k}}}\sum_{x\in \{0,1\}^n,{|x|}=k}~|x\rangle:=|\psi^n_k\rangle.\label{eq.Dicke}
\end{equation}
So finally we arrive at the usual definition of Dicke states. From this simple derivation one could see that the permutation
group ${\mathcal S}_n$  plays a significant role in characterizing spin states.

The crucial strategy in \cite{BE-2019} and \cite{BE-2022} for preparing Dicke states is to introduce the universal unitary operator $U_{n,k}$:
\begin{equation}
U_{n,k} |0^{n-l}1^{l} \rangle =|\psi^n_l\rangle,\quad \forall~~ 0\le l\le k\le n.
\end{equation}
Here the upper bound of the Hamming weights is set to $k$, because their aim is to prepare just $|\psi^n_k\rangle$.
If we want to prepare all the $n$-qubit Dicke states instead, then we could set $k=n$ and consider the even powerful unitary:
\begin{equation}
U_{n,n} |0^{ n-l}1^{ l} \rangle =|\psi^n_l\rangle,\quad \forall~~ 0\le l\le n,\label{eq:Unn}
\end{equation}
In fact, $U_{n,n}$ gives the unitary transformation from a specific computational sub-basis to the maximum-spin states with different spin projections. Of course this is well defined, since both the initial and final states are orthogonal and normalized.
With this definition, we now introduce the explicit constructions given in \cite{BE-2019} and \cite{BE-2022}. Both the algorithms reduce
$U_{n,n}$ recursively into smaller ones. The difference is, \cite{BE-2019} reduces $U_{n,n}$  by one qubit at each step,
while \cite{BE-2022} decomposes $U_{n,n}$ quasi-equally.

\subsection{Split \& Cyclic Shift}
As we said, the strategy of implementing $U_{n,n}$ in \cite{BE-2019} is to reduce it recursively with $n$. This is done based on the following relation:
\begin{equation}
|\psi^n_l\rangle = \sqrt{\frac{n-l}{n}}|\psi^{n-1}_l\rangle \otimes |0\rangle +\sqrt{\frac{l}{n}}|\psi^{n-1}_{l-1}\rangle \otimes |1\rangle,~~ \forall ~~0\le l \le n. \label{eq:Dicke-dec}
\end{equation}
The derivation is almost straightforward, by enumerating the possible states when the last qubit is fixed. Notice that both the two Dicke
states on the right-hand side could be prepared with $U_{n-1,n-1}$, we could first achieve the superposed initial state:
\begin{equation}
\mathrm{SCS}_{n,n-1}|0^{n-l}1^l\rangle=\sqrt{\frac{n-l}{n}}|0^{n-l-1}1^l\rangle \otimes |0\rangle +\sqrt{\frac{l}{n}}|0^{n-l}1^{l-1}\rangle \otimes |1\rangle,~~\forall ~~0\le l \le n. \label{eq:SCS}
\end{equation}
Here SCS is short for ``Split \& Cyclic Shift'', meaning that the new component in the above equation is obtained in this way. Combining Eq. (\ref{eq:Unn}), (\ref{eq:Dicke-dec}), and (\ref{eq:SCS}), we get the following reduction formula:
\begin{equation}
U_{n,n}=(\mathrm{Id} \otimes U_{n-1,n-1})\cdot \mathrm{SCS}_{n,n-1}.\label{eq:Unn2}
\end{equation}
Therefore, $U_{n,n}$ can be implemented by recursively applying $\mathrm{SCS}_{i,i-1}$:
\begin{equation}
U_{n,n}=\prod _{i=2}^n (\mathrm{SCS}_{i,i-1}\otimes \mathrm{Id}^{n-i}).\label{eq:Unn3}
\end{equation}
Notice that the product is taken in increasing order, while the application in the circuit is in decreasing order.

\begin{figure}[h]
\centering
	\includegraphics[width=0.6\textwidth]{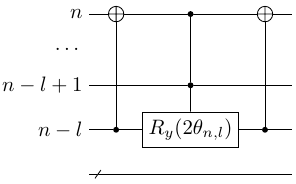}\\
\caption{The circuit $R_{n,l}$ for $2\le l \le n$. Taken from~\cite{BE-2019}.}\label{fig.Rl}
\end{figure}

Now the task has been reduced to constructing the circuit for the individual SCS operator. In \cite{BE-2019} the design of the SCS circuit is described in great detail. Here we only summarize the result. As eq.(\ref{eq:SCS}) shows, the SCS operator is valid universally for arbitrary weight $0\le l\le n$. However, when implementing it we have to treat each weight separately, and then carefully piece together the circuits for all the weights. For $2\le l \le n$, the subroutine of the SCS operator is given by the circuit in Fig.~\ref{fig.Rl}, which we denotes as $R_{n,l}$.

One could see that $R_{n,l}$ is essentially a three-qubit $CCR_y$ rotation together with CNOT gates. The rotation angle is given by
\begin{equation}
\theta_{n,l}=\arccos \sqrt{\frac{l}{n}}.\quad \forall ~2\le l \le n-1. \label{eq:theta}
\end{equation}

When $l=1$, the three-qubit rotation degenerates into a two-qubit rotation, and the corresponding $R_{n,1}$ circuit reads:

\begin{figure}[h]
\centering
	\includegraphics[width=0.6\textwidth]{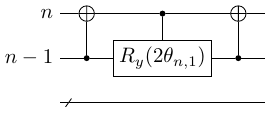}
\caption{The circuit $R_{n,1}$. Taken from~\cite{BE-2019}.}\label{fig.R1}
\end{figure}

And the rotation angle is now $\theta_{n,1}=\arccos \sqrt{\frac{1}{n}}$, which is a special case of eq.~(\ref{eq:theta}).
When we assembly them together, we have to make sure that they do not interfere with each other. This could be guaranteed by
applying $R_{n,l}$ in increasing order of $l$. Thus the complete SCS circuit is given by:
\begin{equation}
\mathrm{SCS}_{n,n-1}=\prod_{l=n-1}^1 \mathrm{Id}^{n-l-1}\otimes R_{n,l}. \label{eq:SCS2}
\end{equation}
An example of the SCS circuit is shown below in Fig.~\ref{fig.SCS}.

\begin{figure}[h]
\centering
	\includegraphics[width=\textwidth]{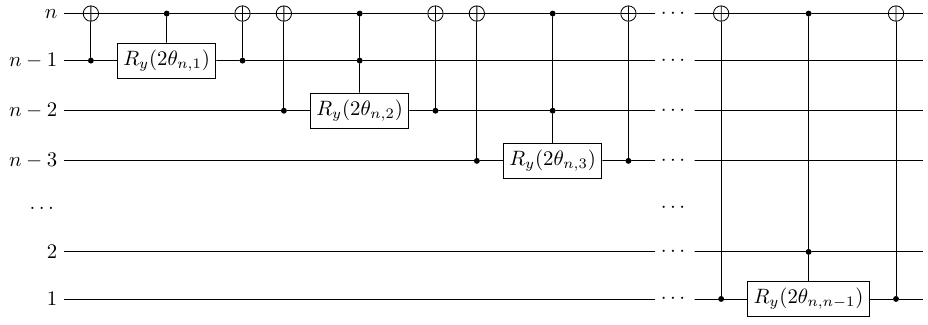}
\caption{The circuit for $\mathrm{SCS}_{n,n-1}$.}\label{fig.SCS}
\end{figure}

Eq. (\ref{eq:Unn3}) and (\ref{eq:SCS2}) now determines the circuit for $U_{n,n}$ completely.
The SCS unitary $\mathrm{SCS}_{n,n-1}$ has an apparent depth ${\mathcal O}(n)$. However, according to the analysis in \cite{BE-2019},
the full unitary $U_{n,n}$ could still be implemented in depth ${\mathcal O}(n)$ utilizing proper parallelization.

\subsection{Weight Distribution Block}

The above strategy is rather straightforward, but not efficient enough. It reduces the full operator $U_{n,n}$ by one qubit at each step, and thus need roughly $n$ steps to finish the job. To improve the efficiency, one would like to reduce the size of $U_{n,n}$ as much as possible at each step. That is to say, we would like it to be decomposed quasi-equally. This is the strategy adopted in \cite{BE-2022}.
To achieve this, another relation for Dicke states is implicitly given in \cite{BE-2022}:
\begin{equation}
|\psi^n_l\rangle=\frac{1}{\sqrt{\tbinom{n}{l}}}\sum_{i=0}^l~\sqrt{\tbinom{m}{i}\tbinom{n-m}{l-i}}~|\psi^m_i\rangle\otimes|\psi^{n-m}_{l-i}\rangle,
~~~\forall ~0\le l\le n.\label{eq:Dicke-dec2}
\end{equation}
The formula is almost obvious from combinatorics, and generalizes (\ref{eq:Dicke-dec}). Now the two Dicke states on the right-hand
side could be prepared with ~$U_{m,m}$~and~$U_{n-m,n-m}$, respectively. Therefore, we should devise the corresponding initial state superposition. This is achieved in \cite{BE-2022} through the so-called weight distribution block~(WDB):
\begin{equation}
\mathrm{WDB}_{m,n-m}|0^{n-l}1^l\rangle=\frac{1}{\sqrt{\tbinom{n}{l}}}\sum_{i=0}^l~~\sqrt{\tbinom{m}{i}\tbinom{n-m}{l-i}}~|0^{m-i}1^i\rangle\otimes|0^{n-m+i-l}1^{l-i}\rangle,~~~\forall ~0\le l\le n.\label{eq:WDB}
\end{equation}
Now combining eqs. (\ref{eq:Unn}), (\ref{eq:Dicke-dec2}) and (\ref{eq:WDB}), we obtain the following recursion relation:
\begin{equation}
U_{n,n}=(U_{m,m} \otimes U_{n-m,n-m})\cdot \mathrm{WDB}_{m,n-m}. \label{eq:U-WDB}
\end{equation}
In principle, this could done all the way until we get the trivial operator $U_{1,1}=\mathrm{Id}$. Then $U_{n,n}$ could be reconstructed by properly assembling all the WDB modules together.

Such a procedure would require us to implement the WDB module for arbitrary weight $0\le l\le n$.
Such a universal WDB module is not realized at the circuit level in \cite{BE-2022}.
Instead, a partial WDB module valid only for a specific weight interval is constructed.
Explicitly, the following weight interval is considered:
\begin{equation}
0 \le l\le k \le \max\{n-m,m\}. \label{eq:weight}
\end{equation}
We may denote the corresponding module as $\mathrm{WDB}^k_{m,n-m}$. And then, the recursive relation (\ref{eq:U-WDB}) becomes:
\begin{equation}
U_{n,k}=(U_{m,k} \otimes U_{n-m,k})\cdot \mathrm{WDB}^k_{m,n-m}. \label{eq:U-WDBk}
\end{equation}
Applying $\mathrm{WDB}^k_{n-m,m}$ recursively, we may reduce the $U_{n,k}$ into several sub-operators $U_{k,k}$.
The previous SCS circuit is then utilized to produce these sub-operators. Such a restricted WDB circuit is also
adopted in later studies, such as~\cite{Vittal-2025}.

Now we review the explicit circuit in \cite{BE-2022} for $\mathrm{WDB}^k_{n-m,m}$. The weight constraint (\ref{eq:weight})
greatly simplifies the designing. Due to eq.(\ref{eq:weight}), all the $|1\rangle$ states in the input would be in one sub-block,
and we just need to distribute part of them to the other sub-block. This could be done by first duplicating some $|1\rangle$ states
to the other sub-block, and then eliminating extra ones in the original sub-block. Still, we should do this for each weight $l$ separately.
This is achieved by taking the one-hot coding of the various initial states. So in total, we have four separated modules: encoding,
duplication, decoding, and elimination. A sketch of the whole circuit is given below in Fig.~\ref{fig.WDB}.

\begin{figure}[h]
\centering
	\includegraphics[width=\textwidth]{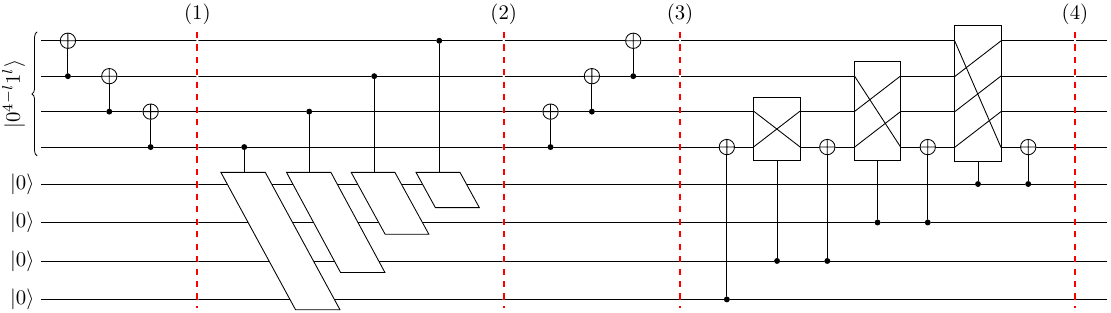}
\caption{A sketch of the $\mathrm{WDB}$ circuit. The details of the second and fourth modules are given below. Taken from \cite{BE-2022}.}\label{fig.WDB}
\end{figure}

Moreover, the duplication module is accomplished by multiple $CCR_y$ gates, extending that of the SCS circuit.
Finally, the elimination module can be easily realized by proper CNOT gates, together with series of Fredkin gates.
All the details are illustrated in Fig.~\ref{fig.dup} and Fig.~\ref{fig.eli} below.

\begin{figure}[h]
\centering
	\includegraphics[width=1.0\textwidth]{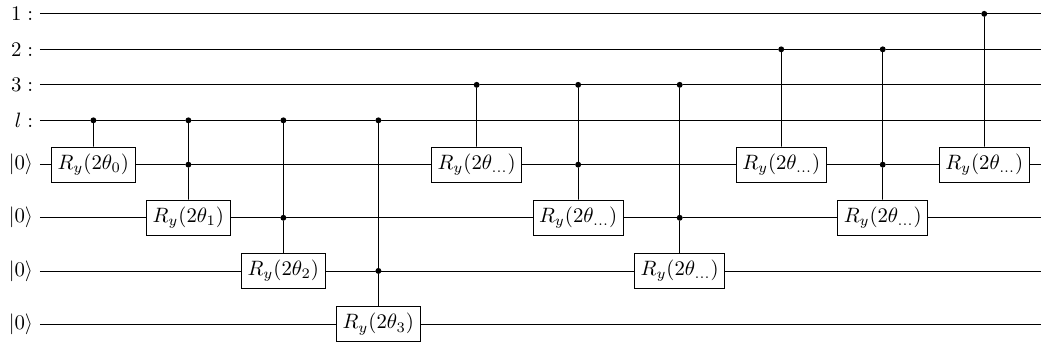}
\caption{An example of the duplication module. Taken from \cite{BE-2022}.}\label{fig.dup}
	\includegraphics[width=1.0\textwidth]{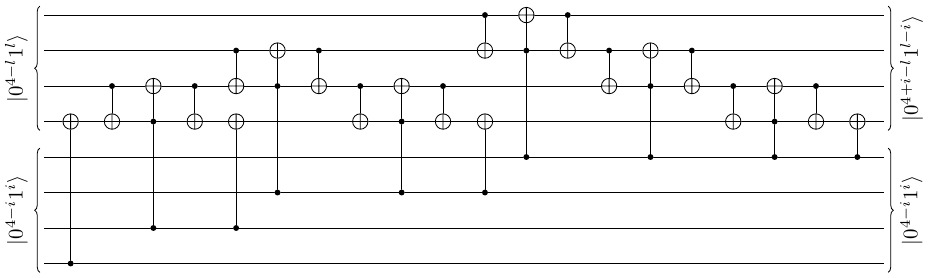}
\caption{An example of the elimination module. Taken from \cite{BE-2022}.}\label{fig.eli}
\end{figure}

The rotation angles in the duplication module are chosen according to the definition (\ref{eq:WDB}),
which we do not repeat here. According to the analysis in \cite{BE-2022}, the above implementation of the restricted unitary
$\mathrm{WDB}^k_{m,n-m}$ has depth ${\mathcal O}(k)$ upon parallelization. In the later sections we will extend the above WDB
circuit from the weight-restricted case to the general case.
%The complexities of the SCS and WDB circuits have been analyzed in detail in \cite{BE-2019} and \cite{BE-2022}, respectively.
 %These will not be our main focus in this paper, and thus are omitted.

\section{Branching Diagram}

Now we start to generalize the two kinds of circuits for Dicke states to arbitrary spin eigenfunctions.
To do this, first we need to specify the spin eigenfunctions explicitly. For Dicke states,
total spin and the spin projection are already enough to completely fix the state. This is certainly
not the case for general spin eigenfunctions, as will be seen soon. A concrete and powerful approach to
achieve this is to utilize group representation theory, as done in \cite{Bacon-2006,Bacon-2007}. However,
the approach seems a little unwieldy for designing practical algorithms. Here we would like to take an intuitive
approach, by tracing the whole forming procedure of the final state. In other words, we use the total spin-coupling
process to characterize the spin eigenfunctions. Since the representations of SU(2) form a symmetric braided category~\cite{Baez-1995},
we could use associativity isomorphisms to rearrange the order of spin couplings. Therefore, we could choose a canonical
scheme to couple the spins. This gives rise to the so-called branching diagram,  which was proposed early in 1930s in the
chemistry community~\cite{VVS-1935}, and further developed in the 1970s~\cite{Pauncz-1977}. We will first review the specification of
spin eigenfunctions in the framework of branching diagram, and then design the proper algorithm to prepare them.
Branching diagram has recently be utilized to prepared specific spin states in~\cite{Sugisaki-2016,Sugisaki-2019}.

\subsection{Branching Diagram}

A very nice review of branching diagram is given in Chapter 2 of \cite{Pauncz-2000}. The underlying idea is that,
we couple the spins one by one in a fixed order. Since coupling a new spin $1/2$ leads to at most two different outcomes,
we could use a branching language to describe it, which is also called ``genealogical construction''. We could further describe
the process in a pictorial way, and this results in the branching diagram. Partial of the diagram for $n\le 6$ is given below in Fig.~\ref{fig.BD}.

\begin{figure}[h]
\centering
	\includegraphics[width=0.75\textwidth]{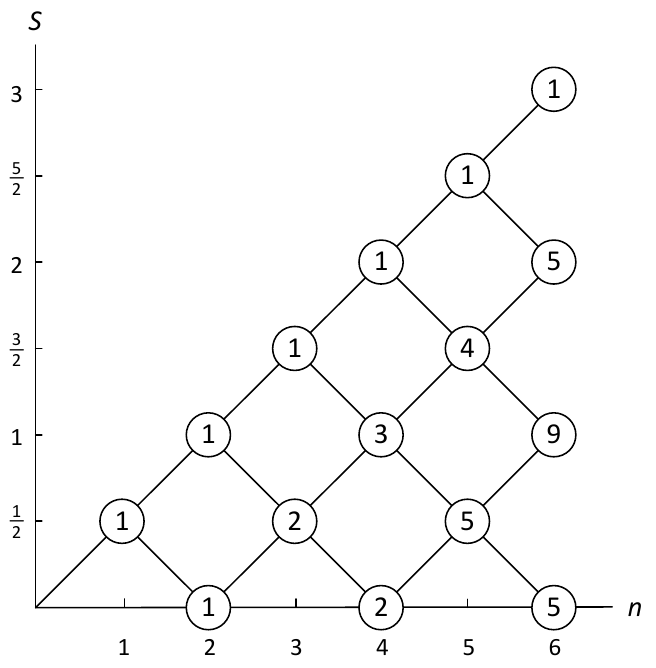}
\caption{Branching Diagram for small $n$. Taken from~\cite{Pauncz-2000}.}\label{fig.BD}
\end{figure}

Let us give some explanations of the diagram. The vertical axis represents the total spin, while the horizontal axis gives the
particle number $n$. The circle with coordinate $(n,S)$ collects the corresponding spin states, and the number inside
the circle denotes the spin degeneracy $f(n,S)$. Explicitly, $f(n,S)$ counts the different branching paths from the origin to
the point $(n,S)$. That is to say, a specific path determines a unique spin state. For example,
the point $(5,1/2)$ has degeneracy $5$, so we have five different paths terminating
at $(5,1/2)$. They are shown explicitly below in Fig.~\ref{fig.BDS}.

\begin{figure}[h]
\centering
	\includegraphics[width=0.8\textwidth]{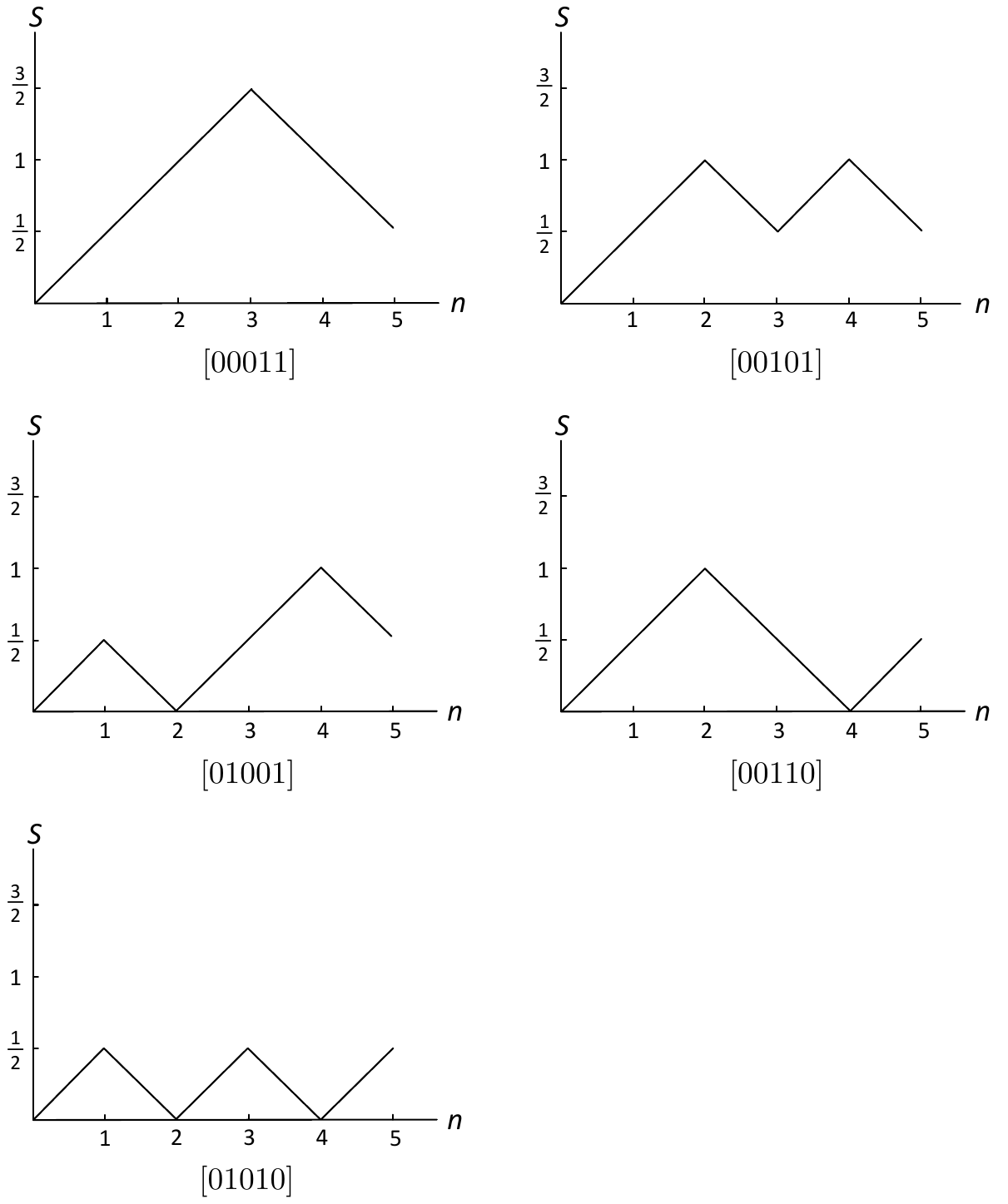}
\caption{Branching paths and branching diagram symbols $b_n$ for $n=5,S=1/2$. Taken from~\cite{Pauncz-2000}.}\label{fig.BDS}
\end{figure}

In the above figure, a unique symbol, called branching diagram symbol, is introduced to represent each branching path~\cite{Pauncz-2000}.
The symbol, denoted as $\vec b_n$, is a $n$-dimensional vector with binary elements. Note that we have slightly changed the
notations from~\cite{Pauncz-2000}. When $\vec b_n[j]=0$, the
spin would be increasing, and decreasing otherwise. We call this the binary representation of the branching diagram symbols.
We could also define a spin vector $\vec \lambda_n$, such that:
\begin{equation}
\vec b_n[i]=0 \Leftrightarrow \vec \lambda_n[i]=+1;\quad \vec b_n[i]=1 \Leftrightarrow \vec \lambda_n[i]=-1,\quad \forall i=1,2,...,n.
\end{equation}
with such a representation, the intermediate spin could be conveniently represented as
\begin{equation}
S_i=\sum_{j=1}^i ~\vec \lambda_n[j]/2,\quad \forall i=1,2,...,n. \label{eq:node-spin}
\end{equation}
And thus the branching diagram symbol $\vec \lambda_n$ should satisfy the positive semi-definite constraints:
\begin{equation}
S_i\ge 0,\quad \forall i=1,2,...,n.
\end{equation}
Since $S_1=\vec \lambda_n[1]/2$, we should always have $\vec \lambda_n[1]=+1$.

When the branching diagram symbol $\vec \lambda_n$ is specified, the corresponding spin eigenfunction would be completely determined.
This is called the branding diagram function~\cite{Pauncz-2000}. For later convenience,
we will use a different notation from~\cite{Pauncz-2000}, and denote it as ~$|S_n,S_n^z,\vec \lambda_n\rangle$.
If every element $\vec \lambda_n[i]$ is chosen to be $1$, then this reproduces the Dicke state. The branching diagram
functions have very nice properties. In particular, for fixed $n$, $|S_n,S_n^z,\vec \lambda_n\rangle$ form a complete
orthonormal basis of spin eigenfunctions. Therefore, we have the following completeness relation:
\begin{equation}
\sum_S (2S+1)~f(n,S)=2^n. \label{eq:ON}
\end{equation}
Explicit expression of $f(n,S)$ and the derivation of the above relation could be found in \cite{Pauncz-2000}.

\subsection{From SCS to CG}
With the branching diagram in hand, the generalization of the SCS circuit for Dicke states to general spin eigenfunctions
would be quite straightforward. For this, we first rewrite the decomposition (\ref{eq:Dicke-dec}) of Dicke states in the
spin language as:
\begin{equation}
|\frac{n}{2},\frac{n}{2}-l\rangle=\sqrt{\frac{n-l}{n}}|\frac{n-1}{2},\frac{n}{2}-l-\frac{1}{2}\rangle\otimes|0\rangle
+\sqrt{\frac{l}{n}}|\frac{n-1}{2},\frac{n}{2}-l+\frac{1}{2}\rangle\otimes|1\rangle.\label{eq:Dicke-dec3}
\end{equation}
As we said, this corresponds to the special branching diagram symbol $\vec \lambda_n=[(+1)^n]$. Now we would like
to generalize the decomposition to arbitrary symbols $\vec \lambda_n$.
Imagine we already arrive at state $|S_i,S_i^z,\vec \lambda_i\rangle$, and we want to further decompose it. Here $\vec \lambda_i$ is
 the shortened vector of $\vec \lambda_n$ by omitting the last $n-i$ elements.
If $\vec \lambda_{i}[i]=+1$, then $S_{i-1}=S_{i}-1/2$, and we have (according to table 10.1 of \cite{Zeng-2013}):
\begin{equation}
|S_{i},S_{i}^z,\vec \lambda_{i}\rangle=\sqrt{\frac{S_i+S_i^z}{2S_i}}|S_i-\frac{1}{2},S_{i}^z-\frac{1}{2},\vec \lambda_{i-1}\rangle\otimes|0\rangle+\sqrt{\frac{S_i-S_i^z}{2S_i}}|S_i-\frac{1}{2},S_{i}^z+\frac{1}{2},\vec \lambda_{i-1}\rangle\otimes|1\rangle.\label{eq:spin-dec1}
\end{equation}
Of course, it includes (\ref{eq:Dicke-dec3}) as a special case. Otherwise, if $\vec \lambda_{i}[i]=-1$, we have $S_{i-1}=S_{i}+1/2$, and
\begin{equation}
|S_{i},S_{i}^z,\vec \lambda_{i}\rangle=-\sqrt{\frac{S_i-S_i^z+1}{2S_i+2}}|S_i+\frac{1}{2},S_{i}^z-\frac{1}{2},\vec \lambda_{i-1}\rangle\otimes|0\rangle+\sqrt{\frac{S_i+S_i^z+1}{2S_i+2}}|S_i+\frac{1}{2},S_{i}^z+\frac{1}{2},\vec \lambda_{i-1}\rangle\otimes|1\rangle.\label{eq:spin-dec2}
\end{equation}
Based on the above decomposition relations, we could now devise the preparation circuit. Following (\ref{eq:Unn}), we define the preparation operator as:
\begin{equation}
U(\vec \lambda_i) |0^{ i-l}1^{ l} \rangle =|S_i,S_i^z=\frac{i}{2}-l,\vec \lambda_i\rangle,\quad \forall~~ 1\le i\le n,\quad \frac{i}{2}-S_i\le l\le \frac{i}{2}+S_i,\label{eq:U-path1}
\end{equation}
$U(\vec \lambda_i)$ is well defined, since the spin $S_i$ appearing on the right-hand side is determined by $\vec \lambda_i$.
Then all the spin states on the the right-hand of both (\ref{eq:spin-dec1}) and (\ref{eq:spin-dec2}) could be prepared universally
with $U(\vec \lambda_{i-1})$. In other words, we could use the inverse of $U(\vec \lambda_{i-1})$ to transform all the spin states back
into computational basis states. So again we may consider the initial superposition circuits instead:
\begin{equation}
\mathrm{CG}^{+1}(\vec \lambda_{i-1})|0^{ i-l}1^{ l} \rangle=\sqrt{\frac{S_i+\frac{i}{2}-l}{2S_i}}|0^{ i-l-1}1^{l} \rangle\otimes|0\rangle+\sqrt{\frac{S_i-\frac{i}{2}+l}{2S_i}}|0^{ i-l}1^{ l-1} \rangle\otimes|1\rangle.\label{eq:CG+}
\end{equation}
\begin{equation}
\mathrm{CG}^{-1}(\vec \lambda_{i-1})|0^{ i-l}1^{ l} \rangle=-\sqrt{\frac{S_i-\frac{i}{2}+l+1}{2S_i+2}}|0^{ i-l-1}1^{l} \rangle\otimes|0\rangle+\sqrt{\frac{S_i+\frac{i}{2}-l+1}{2S_i+2}}|0^{ i-l}1^{ l-1} \rangle\otimes|1\rangle.\label{eq:CG-}
\end{equation}
Here we use the name "CG" to emphasize that all the amplitudes on the right-hand side are now the Clebsch-Gordan~(CG) coefficients in
angular momentum theory. The first circuit corresponds to $\vec \lambda_{i}[i]=+1$, and the second one to $\vec \lambda_{i}[i]=-1$, as shown explicitly
with superscripts. Both circuits are valid for all weights $\frac{i}{2}-S_i\le l\le \frac{i}{2}+S_i$. Since $\vec \lambda_{i}$ has
been provided on the left-hand side, $S_i$ is then fixed, and the right-hand side is well defined. Actually, since the right-hand
side depends only on $S_i$ and not the other spin values, we may replace the variable $\vec \lambda_{i-1}$ on the left hand by $S_{i-1}$.
In other words, the above circuits only depend on the spin values before and after the current coupling, not the entire branching diagram
symbol. Nevertheless, we keep the above notation for formal beauty. We may further succinctly denote the above two operators as
$\mathrm{CG}^{\vec \lambda_{i}[i]}(\vec \lambda_{i-1})$. Then, combining together eqs. (\ref{eq:spin-dec1},\ref{eq:spin-dec2},\ref{eq:CG+},\ref{eq:CG-}) and \ref{eq:U-path1},
we have the recursive relation at each step along the branching path
\begin{equation}
U_{\vec \lambda_i}=(U_{\vec \lambda_{i-1}}\otimes \mathrm{Id})\cdot \mathrm{CG}^{\vec \lambda_{i}[i]}(\vec \lambda_{i-1}).\label{eq:U-path2}
\end{equation}
Taking the composition of all steps, we get the complete construction formula for the objective unitary $U(\vec \lambda_n)$:
\begin{equation}
U(\vec \lambda_n)=\prod_{i=2}^n~\mathrm{CG}^{\vec \lambda_{i}[i]}(\vec \lambda_{i-1})\otimes \mathrm{Id}^{n-i}.\label{eq:U-path2}
\end{equation}
These are simply the generalization of eqs. (\ref{eq:Unn2},\ref{eq:Unn3}).

\subsection{SCS-CG Circuits}

We still need to construct explicitly the quantum circuits for $\mathrm{CG}^{\vec \lambda_{i}[i]}(\vec \lambda_{i-1})$.
Here we could take a shortcut by generalizing the previous SCS circuit directly. If we compare (\ref{eq:CG+}) and (\ref{eq:CG-}) with
(\ref{eq:SCS}) carefully, we immediately find that all of them possess the same structure. The only differences between them
are the coefficients. Therefore, we could utilize the SCS circuit, and generalize it by adjusting the parameters properly.
First, for $\vec \lambda_{i}[i]=+1$, we define the angles according to (\ref{eq:CG+}):
\begin{equation}
\theta_l^{+1}(\vec \lambda_{i-1})=\arccos \sqrt{\frac{S_i-\frac{i}{2}+l}{2S_i}},\quad \forall~~ \frac{i}{2}-S_i\le l\le \frac{i}{2}+S_i. \label{eq:theta1}
\end{equation}
And similarly, for $\vec \lambda_{i}[i]=-1$, we define the angles based on (\ref{eq:CG-}):
\begin{equation}
\theta_l^{-1}(\vec \lambda_{i-1})=-\arccos \sqrt{\frac{S_i+\frac{i}{2}-l+1}{2S_i+2}}, \quad \forall~~ \frac{i}{2}-S_i\le l\le \frac{i}{2}+S_i. \label{eq:theta2}
\end{equation}
Again we write them succinctly as $\theta_l^{\vec \lambda_{i}[i]}(\vec \lambda_{i-1})$. Now replacing the angles $\theta_{i,l}$ in
the previous $R_{i,l}$ circuits by $\theta_l^{\vec \lambda_{i}[i]}(\vec \lambda_{i-1})$, we obtain new circuits
$R_l^{\vec \lambda_{i}[i]}(\vec \lambda_{i-1})$. Then the CG unitary could be achieved by assembling these sub-circuits properly as:
\begin{equation}
\mathrm{CG}^{\vec \lambda_{i}[i]}(\vec \lambda_{i-1})=\prod_{l=\frac{i}{2}+S_i-1}^{\frac{i}{2}-S_i+1} \mathrm{Id}^{n-l-1}\otimes R_l^{\vec \lambda_{i}[i]}(\vec \lambda_{i-1}). \label{eq:SCS-CG}
\end{equation}
We may call it a ``SCS-CG'' circuit. Essentially this simply generalizes (\ref{eq:SCS2}) by adjusting the weight range and
the corresponding angles. Substituting (\ref{eq:SCS-CG}) into (\ref{eq:U-path2}), we get the full circuit
for $U(\vec \lambda_n)$, as desired. Since the SCS-CG circuit has exactly the same structure as the original SCS circuit, the full unitary
$U(\vec \lambda_n)$ could also be implemented in depth ${\mathcal O}(n)$ with proper parallelization.

\section{Spin Tree}

In the previous section we manage to prepare all the branching diagram functions $|S_n,S_n^z,\vec \lambda_n\rangle$ described by branching paths.
They constitute a canonical kind of spin eigenfunctions, and form a complete orthonormal basis,
as we mentioned before. However, such a successive coupling scheme may not describe the true coupling procedure. In other words,
in practice we may not always coupling the spins in a successive way. In this section we try to generalize the successive coupling
scheme to more realistic situations. Historically, the branching diagram framework is also called ``genealogical construction'', as we mentioned before.
And for a genealogical description, the natural language would be the binary tree. So we will try to generalize the branching paths to
binary trees.

\subsection{From Paths to Trees}

First we would like to demonstrate that a branching path is secretly a binary tree in a specific way. To show this,
we need to modify the branching path to include explicitly the information of all input spins. Let us focus on a given branching path.
We omit the first trivial edge from point $0$ to $1$. Then we get a path with $n$ vertices and $n-1$ edges.
 Now we transform it into a specific binary tree with the following steps:
\begin{framed}
\noindent \emph{\hspace*{1em} 1. Take the line graph of the path. Concretely, we turn each vertex into a new edge, each edge into a new vertex,
and keep the correlation relationship. The initial vertex is turned into a leaf edge, and the terminal vertex is turned into the
root edge;\\
\hspace*{1em} 2. Add a new leaf edge at each new vertex;\\
\hspace*{1em} 3. For each new vertex, if the original edge is ascending, label it with ``+1''; otherwise, label it with ``-1'';\\
\hspace*{1em} 4. Delete all degree-1 external vertices. In doing so, we make all leaf edges and the root edge dangling.}
\end{framed}

Obviously, the resulting tree has the following properties:
\begin{framed}
\noindent \emph{
\hspace*{1em} 1. It contains $n$ leaf edges, $1$ root edge;\\
\hspace*{1em} 2. It contains $n-1$ internal vertices, all with degree 3;\\
\hspace*{1em} 3. Each vertex is incident with at least one leaf edge;\\
\hspace*{1em} 4. The labels on the vertices together with $\vec \lambda_n[1]\equiv +1$ determine the branching vector $\vec \lambda_n$.}
\end{framed}

In short, it is a planted proper binary (rooted) tree with all degree-1 external vertices removed,
and with ``+1/-1'' labels on the internal vertices. In this way, we can always generate
such a special binary tree from a given branching path. The reverse could also been done.
Given such a binary tree, the labels on the vertices together with $\vec \lambda_n[1]\equiv +1$ generate a branching path vector.
If we further impose the following positive semidefinite conditions
\begin{equation}
S_i=\sum_{j=1}^i ~\vec \lambda_n[j]/2\ge 0, ~~~\forall ~1\le i \le n,
\end{equation}
the branching path would be legal. An example of the path-tree correspondence is given below in Fig.~\ref{fig.path-tree}.

\begin{figure}[h]
\centering
	\includegraphics[width=0.8\textwidth]{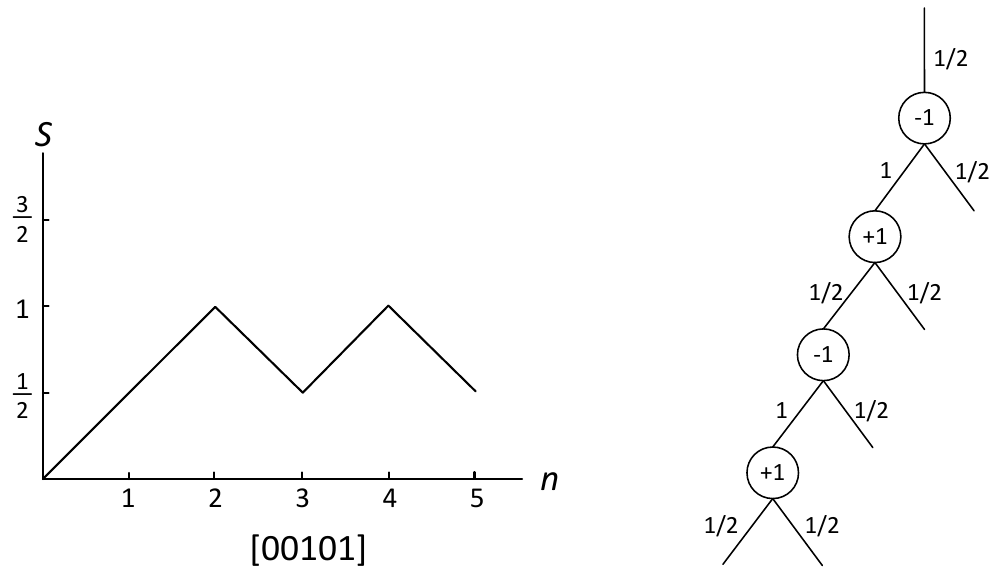}
\caption{An example of the branching path-tree correspondence.}\label{fig.path-tree}
\end{figure}

\subsection{Spin Tree}

Now we attempt to generalize the above binary tree. We keep the underlying graphic structure of the tree,
namely a planted proper binary tree. But we relax the condition that each vertex is incident with at least one
leaf edge. Once we make this relaxation, the ``+1/-1'' label on the vertex is not enough to fix the coupling process.
Instead, we have to explicitly specify all the spins on the intermediate
edges. In this way, we are led to a certain kind of ``spin tree''. Below we give its formal definition:
\begin{framed}
\noindent \textbf{Definition: Spin Tree $ST^n:=\{n,T^n,{\mathcal S}\}$:} \\
\emph{\hspace*{1em} i) $T^n$ is a planted proper binary (rooted) tree with $n$ leaf vertices, and with all degree-1 vertices removed, such that:\\
\hspace*{2em} 1. It contains a dangling root edge $r$;\\
\hspace*{2em} 2. It contains $n$ dangling leaf edges;\\
\hspace*{2em} 3. It contains $n-1$ internal vertices, all of degree 3.\\
\hspace*{1em} ii) ${\mathcal S}: E(T^n) \to \{0\}\cup \mathbb{N}/2$ is an assignment of spins on edges, such that:\\
\hspace*{2em} 1. The spin on each leaf edge, or leaf spin, is fixed to be 1/2;\\
\hspace*{2em} 2. When $n$ is even, the spin on the root edge, or root spin, takes value from
$\{0,1,2,...,n/2\}$; when $n$ is odd, the root spin takes value from $\{1/2,3/2,...,n/2\}$;\\
\hspace*{2em} 3. On each (internal) vertex, the three incident spins satisfy the triangular rule and the integer rule:
\begin{equation}
J_1\le J_2+J_3,\quad J_2\le J_1+J_3,\quad J_3\le J_1+J_2,\quad J_1+J_2+J_3\in \mathbb{Z}.\label{eq:Triangle-Integer}
\end{equation}}
\end{framed}
In fact, condition 1 and 3 of the spin assignment together imply 2. However, since condition 2 is very important
for the circuit designing later, we still keep it here. In the previous subsection,
we have shown that a branching path is equivalent to a specific binary tree with a unique branch.
We could further rewrite it in the spin language, and call it a ``branching spin tree'', or just branching tree for short.

Readers familiar with Penrose's spin network~\cite{Penrose-1971} may have noticed that the above
spin trees bear remarkable similarity to spin networks. Indeed,  when conceiving the concept of spin tree we already
have spin network in mind. However, there is a significant
difference between them. Spin networks do not possess ``time'', and thus describe quantum states. Our spin
trees actually describe quantum processes, the forming processes of different spin states. According to the philosophy of~\cite{JB-2010},
this helps us to design the corresponding circuits.

\subsection{From WDB to CG}

Comparing to branching diagram, spin trees characterize general spin coupling processes. How could we prepare the spin state
resulting from a given spin tree? The answer is simple: we could achieve this with the general form of the WDB circuits.
While the SCS circuits naturally suit the branching diagram framework, the WDB circuits naturally suit the spin tree formalism.
Such a relation between the WDB circuits and binary trees is explicitly shown recently in~\cite{Vittal-2025}.
Namely, the recursive relation (\ref{eq:U-WDB}) based on WDB could be identified as the decomposition on a vertex of the spin tree.
However, to make this statement concrete, we need to rewrite the whole formalism for the WDB circuits.

Recall that, the underlying principle for the WDB circuits is the decomposition relation (\ref{eq:Dicke-dec2}) of the Dicke state,
which we rewrite below:
\begin{equation}
|\psi^n_l\rangle=\frac{1}{\sqrt{\tbinom{n}{l}}}\sum_{i=0}^l~\sqrt{\tbinom{m}{i}\tbinom{n-m}{l-i}}~|\psi^m_i\rangle\otimes|\psi^{n-m}_{l-i}\rangle, \forall~ 0\le l\le n.\label{eq:Dicke-dec4}
\end{equation}
Since Dicke states can be identified with maximum-spin states, we could rewrite the above relation in spin language:
\begin{equation}
|\frac{n}{2},\frac{n}{2}-l\rangle=\sum_{i=0}^l~A_{m,n-m}^{i,l-i}|\frac{m}{2},\frac{m}{2}-i\rangle\otimes|\frac{n-m}{2},\frac{n-m}{2}-l+i\rangle.\label{eq:Dicke-dec5}
\end{equation}
Here the coefficients $A_{m,n-m}^{i,l-i}$ is the corresponding CG coefficients:
\begin{equation}
A_{m,n-m}^{i,l-i}=\langle \frac{m}{2},\frac{m}{2}-i;\frac{n-m}{2},\frac{n-m}{2}-l+i|\frac{n}{2},\frac{n}{2}-l\rangle.\label{eq.Amn0}
\end{equation}
It is not difficult to verify that, eq.~(\ref{eq:Dicke-dec5}) indeed reproduces (\ref{eq:Dicke-dec4}) once the CG coefficients
are explicitly derived. Now we could utilize the powerful formalism of angular momentum theory to generalize (\ref{eq:Dicke-dec5})
to arbitrary spin eigenstates. But just as in the situation of branching paths, we first need to characterize the spin states properly.
With the notion of spin tree, we could do this straightforwardly.

For a spin tree $ST^n$ with root spin $S$,  we denote the outcome spin state as $|S,S^z,ST^n\rangle$.
Now we delete the root edge and the vertex incident with it. We get two spin subtree $ST_L^m$ and $ST_R^{n-m}$,
with root spins $S_L$ and~$S_R$ respectively. According to the angular momentum coupling rules, we have:
\begin{equation}
|S,S^z,ST^n\rangle=\sum_{S_L^z+S_R^z=S^z}\langle S_L,S_L^z;S_R,S_R^z|S,S^z\rangle~|S_L,S_L^z,ST_L^m\rangle\otimes |S_R,S_R^z,ST_R^{n-m}\rangle. \label{eq:Spin-Tree-Dec}
\end{equation}
Here the CG coefficients have been explicitly shown.
%It should emphasized that these CG coefficients depend only on the spins and spin projecttions
%involved, not on the whole tree structure.

The above decomposition (\ref{eq:Spin-Tree-Dec}) generalizes that of Dicke states (\ref{eq:Dicke-dec5}) dramatically.  This would allow
us to prepare these spin states recursively. To do this, first we define the preparation unitary by analogy with (\ref{eq:Unn}) as:
\begin{equation}
U(ST^n) |0^{ n-l}1^{ l} \rangle =|S,\frac{n}{2}-l,ST^n\rangle,\quad \forall~~ \frac{n}{2}-S\le l\le \frac{n}{2}+S.\label{eq:U-tree}
\end{equation}
Since the states on either side are orthonormal, the operator is well defined. Then we define the
superposition operator based on the decomposition (\ref{eq:Spin-Tree-Dec}):
\begin{eqnarray}
\mathrm{CG}_{m,n-m}(S_L,S_R;S)|0^{n-l}1^l\rangle&=&\sum_{i=0}^l~\langle S_L,\frac{m}{2}-i;S_R,\frac{n-m}{2}-l+i|S,\frac{n}{2}-l\rangle\nonumber\\
&& ~ |0^{m-i}1^i\rangle\otimes|0^{n-m+i-l}1^{l-i}\rangle,\quad \forall~~ \frac{n}{2}-S\le l\le \frac{n}{2}+S.\label{eq:CG-Tree}
\end{eqnarray}
%Notice that this circuit only depends on the local information of the vertex incident with the root edge.
Combining eqs.(\ref{eq:Spin-Tree-Dec},
\ref{eq:U-tree},\ref{eq:CG-Tree}), we obtain the recursive relation for the preparation operator:
\begin{equation}
U(ST^n)=[U(ST^m_L)\otimes U(ST^{n-m}_R)]\cdot \mathrm{CG}_{m,n-m}(S_L,S_R;S). \label{eq:U-CG}
\end{equation}
Repeating this procedure recursively, we could eventually obtain the whole preparation unitary $U(ST^n)$. But there is one problem left:
how to implement the CG circuit (\ref{eq:CG-Tree}) in this case?

\subsection{WDB-CG Circuits}

Comparing the CG unitary (\ref{eq:CG-Tree}) with the original WDB unitary (\ref{eq:WDB}), we can see that they have exactly the same structure.
The only differences come from those coefficients. So once we successfully implement the WDB operator, we could immediately adapt it to
the CG case. However, as we mentioned before, the full WDB unitary is not achieved in \cite{BE-2022}. Only the truncated
 $\mathrm{WDB}^k_{m,n-m}$ is realized concretely, which is valid for the restricted weight range:
\begin{equation}
0 \le l\le k \le \max\{n-m,m\}. \label{eq:weight2}
\end{equation}
We now generalize the circuit for $\mathrm{WDB}^k_{m,n-m}$ to the full unitary $\mathrm{WDB}_{m,n-m}$,
valid for the whole range $0\le l\le n$. Our strategy would be as follows. When $l\le n-m$,
we borrow the original WDB circuit, and transfer $|1\rangle$ states from one block to the other block. When $n-m<l\le n$, we will have $n-l< m$,
 which counts the number of $|0\rangle$ states. We then construct a negative-logic WDB circuit to transfer $|0\rangle$ states the other way around. Thus
we could devise the whole circuit with the following steps:
\begin{framed}
\noindent\emph{
\hspace*{1em} 1. Introduce an auxiliary qubit $q_0$, and use its $|0\rangle$ and $|1\rangle$ states to indicate $l\le n-m$ and $n-m<l\le n$
respectively;\\
\hspace*{1em} 2. Apply $\mathrm{WDB}^{n-m}_{m,n-m}$ when $q_0$ is in the $|0\rangle$ state;\\
\hspace*{1em} 3. Apply a negative logic unitary $\overline{\mathrm{WDB}}^{m}_{m,n-m}$ when  $q_0$ is in the $|1\rangle$ state;\\
\hspace*{1em} 4. Perform the uncomputation to restore $q_0$.}
\end{framed}

Let us give some details. Step 1 is easy to implement with a single CNOT gate. Step 2 could be directly borrowed from \cite{BE-2022}.
Step 3 could then be obtained by properly reversing the logic of the original WDB circuit in \cite{BE-2022}. Step 4 is a little tricky,
since all the $|1\rangle$ states and $|0\rangle$ states have been redistributed. But we could still manage to properly estimate the total Hamming
weight of the output state, and then use it to restore $q_0$. Concretely, we need a series of $C\bar C C X$ gates, together its
degenerate versions~($CCX$ or $\bar CCX$ gates) at the beginning and the end. Once we get the full unitary $\mathrm{WDB}_{m,n-m}$, we could immediately obtain the analogous CG unitary
$\mathrm{CG}_{m,n-m}(S_L,S_R;S)$ by properly adjusting the rotation angles. The general strategy for choosing these angles is described in detail
in Appendix.\ref{App.angle}.

\begin{figure}[h]
\centering
	\includegraphics[width=\textwidth]{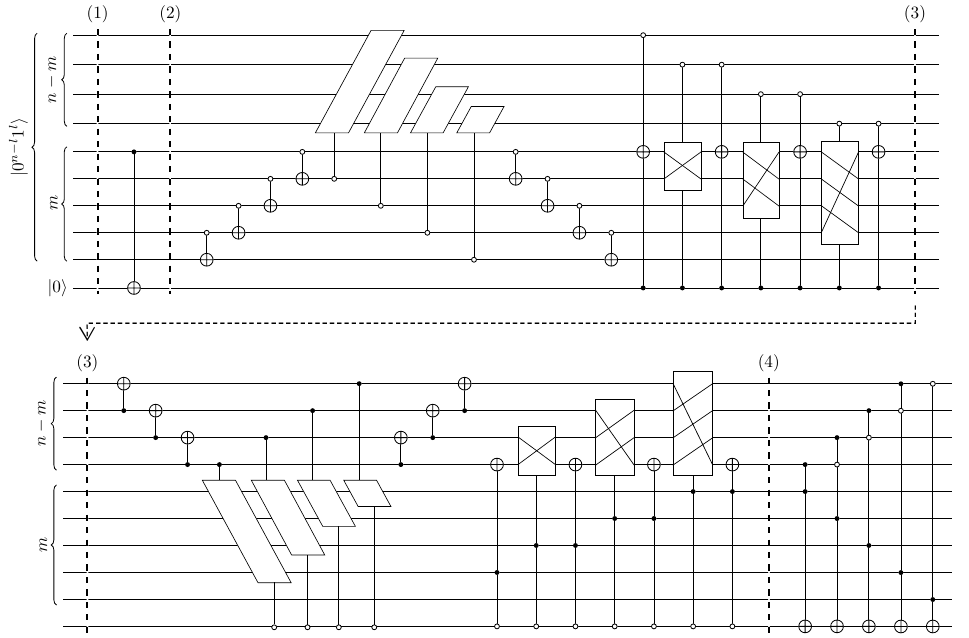}
\caption{An example of the WDB-CG circuit, with the four steps indicated.}\label{fig.WDB-CG}
\end{figure}

The final circuit is called a ``WDB-CG'' circuit. Note that when we go from Dicke states to the general spin states, the weight
range changes accordingly. So we should modify the content of the corresponding WDB blocks in the circuit accordingly.
An example of the WDB-CG circuit is given in Fig.~\ref{fig.WDB-CG}.

We should emphasize that, the WDB-CG circuit designed this way may not be optimal. Utilizing parallelization, the original truncated WDB
circuit could be implemented in depth ${\mathcal O}(k)$. Such a parallelizability is clearly lost due to the presence of the
auxiliary qubit. As a result, our circuit can only be implemented in depth ${\mathcal O}(n^2)$. Could we improve such a circuit to at least
linear depth? And, could we get rid of the auxiliary qubit? The two questions could be related. We would like to investigate them in the future.

\section{Classical Verification}

In this section we make some verifications on the two preparation circuits. First we show that both of them correctly prepare the
Lieb-Mattis states, which we will introduce shortly. Then, we collect and develop the corresponding classical algorithms to make further
verifications.

\subsection{Lieb-Mattis states}

It is easy to show, Dicke states are the maximum-energy states of Heisenberg model on complete graphs. When the model is defined on
complete bipartite graphs, the corresponding ground states are found by Lieb and Mattis in 60's~\cite{LM-1962}, and thus named
Lieb-Mattis~(LM) states. A detailed derivation of these results are given recently in the appendix of~\cite{CM-2013}. We summarize
these results below, and then use the previous two circuits to prepare them explicitly.

The (anti-ferromagnetic)-Heisenberg model is defined by the following Hamiltonian:
\begin{equation}
H^\textrm{H}:=\sum_{ij\in E} (X_iX_j+Y_iY_j+Z_iZ_j).\label{eq:Heisenberg}
\end{equation}
The summation is performed over the edge set of the underlying graph. It turns out convenient to formulate it in a different way,
as the so-called quantum maximum cut (QMC) problem~\cite{AGM-2020}. The corresponding Hamiltonian is defined as
\begin{equation}
H^\textrm{QMC}:=\frac{1}{2}\sum_{ij\in E} (I_iI_j-X_iX_j-Y_iY_j-Z_iZ_j).\label{eq:QMC}
\end{equation}
It has the nice property that, each term is positive semi-definite. Therefore, all the eigenvalues are non-negative. Moreover,
using the SWAP gate (\ref{eq:SWAP}), we may rewrite the Hamiltonian as:
\begin{equation}
H^\textrm{QMC}=\sum_{ij\in E} (I_iI_j-F_{ij}).
\end{equation}

Now if the underlying graph is complete, we could use eq.~(\ref{eq:Spin-swap}) to simplify it into
\begin{equation}
H^\textrm{QMC}(K_n)=\frac{n}{2}(\frac{n}{2}+1)-{\vec S}^2.\label{eq:QMC-spectra}
\end{equation}
So the state with maximum cut is given by those with minimum spin. Related results are given recently in \cite{APS-2025}.
A simple way to achieve these states is to pair spins anti-parallelly two by two. The situation would be a little more difficult
when we go to complete bipartite graphs. Still, the states with maximum energy/cut could be obtained in a similar way. We adopt the framework
of \cite{Watts-2023}, namely the clique-decomposition of graphs, to do the derivation. For complete bipartite graph $K_{m,n}$~($m\ge n$), the decomposition is simple:
\begin{equation}
K_{m,n}=(K_m\cup K_n)^c.
\end{equation}
Then the Hamiltonian could be written as
\begin{equation}
H^\mathrm{QMC}(K_{m,n})=H^\mathrm{QMC}(K_{m+n})-H^\mathrm{QMC}(K_{m})-H^\mathrm{QMC}(K_{n}).
\end{equation}
Now each term could be simplified with (\ref{eq:QMC-spectra}). Thus we obtain
\begin{eqnarray}
H^\mathrm{QMC}(K_{m,n})&=&\frac{m+n}{2}(\frac{m+n}{2}+1)-\frac{m}{2}(\frac{m}{2}+1)-\frac{n}{2}(\frac{n}{2}+1)\nonumber\\
&&-{\vec S_{m+n}}^2+{\vec S_m}^2+{\vec S_n} ^2.
\end{eqnarray}
Here $\vec S_m, \vec S_n$ and $\vec S_{m+n}$ denote the spin operator of $K_m$, $K_n$ and $K_{m+n}$ respectively, so $\vec S_{m+n}=\vec S_m+\vec S_n$.
According to the angular momentum theory,
we could simultaneously determine the square moduli of the total spin and the individual sub-spins. Therefore all the eigenvalues
and eigenstates
could be obtained by specifying the spins and the corresponding spin states. In particular, the maximum cut is achieved
by maximizing sub-spins and minimizing the total spin. Explicitly, by taking
\begin{equation}
S_m=\frac{m}{2},\,\,S_n=\frac{n}{2},\,\,S_{m+n}=\frac{m-n}{2},
\end{equation}
we obtain the maximum cut value
\begin{equation}
\lambda_\mathrm{max}(H^\mathrm{QMC}(K_{m,n}))=mn+n.
\end{equation}
As shown in \cite{APS-2025}, it saturates the matching upper bound~\cite{APS-2025,BBKL-2026}. The corresponding state
could be constructed as follows. First we construct Dicke states for each clique, then we pair the two Dicke states anti-parallelly
to achieve minimum total spin. These are the so-called LM states. The explicit form of the LM states could be written as
\begin{equation}
|\phi_{m,n;l}^{\textrm{LM}}\rangle:=\sum_{k=0}^l A_{m,n}^{k,l-k} |\psi_k^m\rangle |\psi_{l-k}^n\rangle,~~\forall ~~n\le l\le m, \label{eq.ggLM}
\end{equation}
where the parameter $l$ denotes the Hamming weight of the final state, and the CG coefficients are
\begin{equation}
A_{m,n}^{k,l-k}=\langle \frac{m}{2},\frac{m}{2}-k;\frac{n}{2},\frac{n}{2}+k-l|\frac{m-n}{2},\frac{m+n}{2}-l\rangle,~~\forall ~~n\le l\le m. \label{eq.Amnl-1}
\end{equation}
Since the total spin is $S=(m-n)/2$, the constraint for $l$ ensures that
\begin{equation}
-S\le S^z=\frac{m+n}{2}-l\le S.
\end{equation}
When $m=n$, $l$ is forced to be $n$ too, and the final state is of spin 0. Then the CG coefficients simplify to~(\cite{Zeng-2013}, section 10.4)
\begin{equation}
A_{n,n}^{k,n-k}=\langle \frac{n}{2},\frac{n}{2}-k;\frac{n}{2},k-\frac{n}{2}|0,0\rangle=\frac{(-1)^k}{\sqrt{n+1}}.
\end{equation}
And the LM states simplify accordingly:
\begin{equation}
|\phi_{n,n}^{\textrm{LM}}\rangle:=\frac{1}{\sqrt{n+1}}\sum_{k=0}^n(-1)^k |\psi_k^n\rangle |\psi_{n-k}^n\rangle.\label{eq:LM0}
\end{equation}
Perhaps this is the most studied case.
%We intentionally repeat all these elementary derivations, so that colleagues from the computer science
%side could get familiar with them easily.

The special LM states (\ref{eq:LM0}) have recently been prepared in \cite{Marti-2025}, by directly entangling two sets of Dicke states.
Here we
use our new circuits to prepare the general LM states (\ref{eq.ggLM}). According to the above derivation, we could design a natural
spin tree. First we construct two branching subtrees to achieve the Dicke states, then couple them anti-parallelly.  Such a spin tree
would be very special. We may attempt to utilize
association isomorphisms to express them as combinations of branching trees. It turns out that, in this case the association isomorphisms
degenerate to identities, or the association laws.  As a result, we get a unique branching tree, namely the branching diagram symbol
$\vec \lambda_{m+n}=[(+1)^m,(-1)^n]$. Obviously such a degeneration also occurs on branching trees. So we will be free to adjust the branching
sub-trees into an arbitrary coupling scheme. With all these preparations, we could then use either the SCS-CG circuits or the WDB-CG circuits
to produce the general LM states.

\begin{figure}[h]
\centering
	\includegraphics[width=0.6\textwidth]{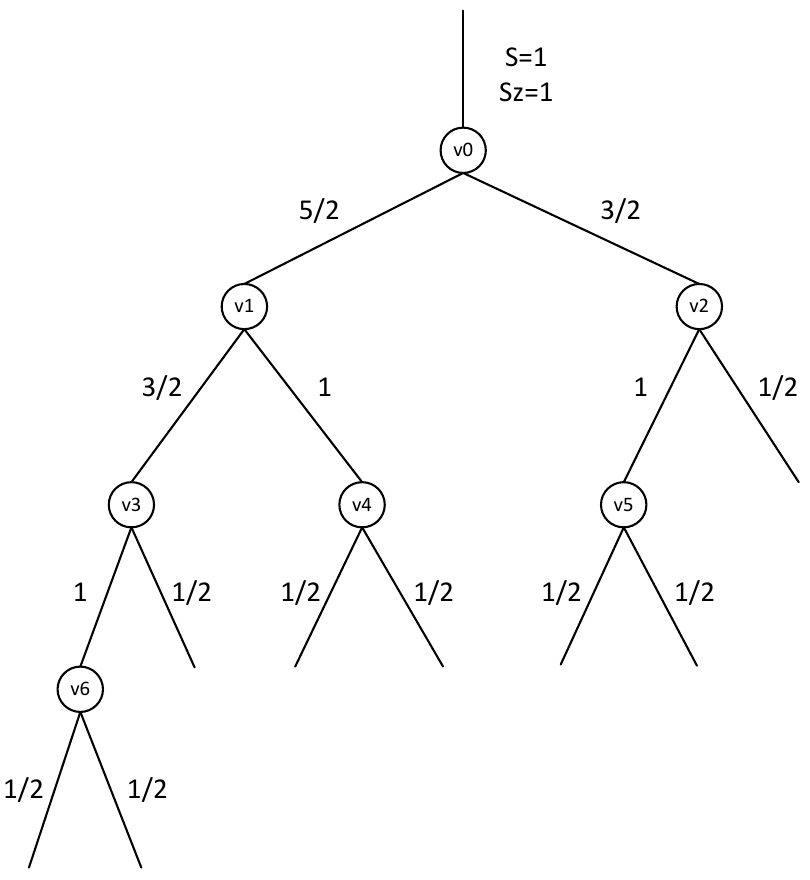}
\caption{A spin tree for a general LM state $|\phi_{5,3;3}^{\textrm{LM}}\rangle$.}\label{fig.LM}
\end{figure}

For example, for $m=5$, $n=3$, $l=3$, we could generate the LM state from the specific spin tree in Fig.~\ref{fig.LM}.

Alternatively, we could characterize it with the branching diagram symbol $\vec \lambda_{8}=[(+1)^5,(-1)^3]$. We could then generate
the corresponding WDB-CG circuits and SCS-CG circuits, and assemble them respectively to produce the required state. The whole circuits
are a little complicated, and will not be shown explicitly. Both of them reproduce the desired state $|\phi_{5,3;3}^{\textrm{LM}}\rangle$
defined in (\ref{eq.ggLM}).

\subsection{General Framework}

While the full expression of the LM states (\ref{eq.ggLM}) could be easily written down in a compact way, this would not be the case for the
general spin states. In general, the classical derivation of a specific spin state could be rather tedious. If we do this forwardly
along a given coupling procedure, a large amount of intermediate information needs to be stored, and the calculation soon slows down.
This could be
greatly improved by taking a backward procedure, starting with the target state. In fact, both the SCS and WDB circuits for
Dicke states are devised in such a backward way. The adaption from the quantum situation to the classical counterpart would be
straightforward. Even so, such a classical construction would still not be efficient enough in general. Nevertheless, these improved classical
algorithms could be used to verify our quantum circuits in small systems.

\subsubsection{Classical Evaluation on Branching Diagram}

We start with the branching diagram. First we ask such a question: for a given pair $(n,S)$, how to generate all the $f(n,S)$ branching diagram
symbols? (Exercise 2.3 of~\cite{Pauncz-2000}). If we follow literally the definition of branching diagram, we would start from the origin,
take each branching step by step, prune those reaching negative spins and keep all the others. Such a forward procedure would produce
all the branching diagram symbols for all spins and all particle numbers. For a given pair $(n,S)$, we could do this backwardly.
Starting from $S_n$, we list all the possible values of the intermediate spins $S_i$ recursively.
This is done according to the following two rules, which have already been mentioned before:
\begin{framed}
\noindent \emph{\hspace*{1em}  1. Global rule: $0\le S_i\le \frac{i}{2}$;\\
\hspace*{1em}  2. Local (branching) rule: $S_{i-1}=S_i\pm 1/2$.}
\end{framed}
In practise, we could at each step invoke the local branching rule first, and then prune those paths that violate the global rule.
With all the spin values on a specific path, we could immediately get the symbol $\vec \lambda_n$ from (\ref{eq:node-spin}). In this way,
we could exactly enumerate all $f(n,S)$  branching paths, no more and no less.

Now fix a branching diagram symbol $\vec \lambda_n$, how could we reproduce the spin state $|S_n,S_n^z,\vec \lambda_n\rangle$
in a classical way? In other words, we want to enumerate all the components appearing in it, and further obtain all the
coefficients/amplitudes. This problem has been thoroughly solved in the quantum chemistry literature. See chapter 2 of
\cite{Pauncz-2000} for the details. We briefly summarize these results below.

%We absorb some of the notations of \cite{Pauncz-2000}
%into our formalism, to make the presentation in accordance with \cite{Pauncz-2000}.

Recall that our branching diagram symbol can either be represented by $\vec \lambda_n\in \{+1,-1\}^n$, or by $\vec b_n\in \{0,1\}^n$.
They could be converted through the relations:
\begin{equation}
\vec \lambda_n[j]=+1 \Leftrightarrow \vec b_n[j]=0;\quad \vec \lambda_n[j]=-1 \Leftrightarrow \vec b_n[j]=1.
\end{equation}
Now a basis state in the computational basis is born with a binary vector $\vec p_n$. This is the so-called ``primitive spin function'' in
\cite{Pauncz-2000}. We could also convert it into a spin representation $\vec \rho_n$ through:
\begin{equation}
\vec p_n[j]=0 \Leftrightarrow \vec \rho_n[j]=+1;\quad \vec p_n[j]=1 \Leftrightarrow \vec \rho_n[j]=-1.
\end{equation}
Then the $j$-particle has spin projection $\rho_n[j]/2$. Summing them up, we obtain:
\begin{equation}
S^z_i=\sum_{j=1}^i ~\vec \rho_n[j]/2,\quad  \forall ~~1\le i \le n.\label{eq:edge-spin3}
\end{equation}
We could plot all the pairs $(i,S^z_i)$ on the plane, with $i$ the horizontal coordinate, and $S^z_i$ the vertical coordinate, and connect successive nodes by edges. This gives rise to the so-called ``path diagram''~\cite{Pauncz-2000}. To avoid confusion, we call them primitive path diagrams,
or primitive paths for short. The primitive path will be ascending when $\vec \rho_n[j]=+1$, and descending otherwise. Thus $\vec \rho_n$ is termed ``path diagram symbol''. Two examples of path diagrams are given below in Fig.~
\ref{fig.path-diagram}.

\begin{figure}[h]
\centering
	\includegraphics[width=0.85\textwidth]{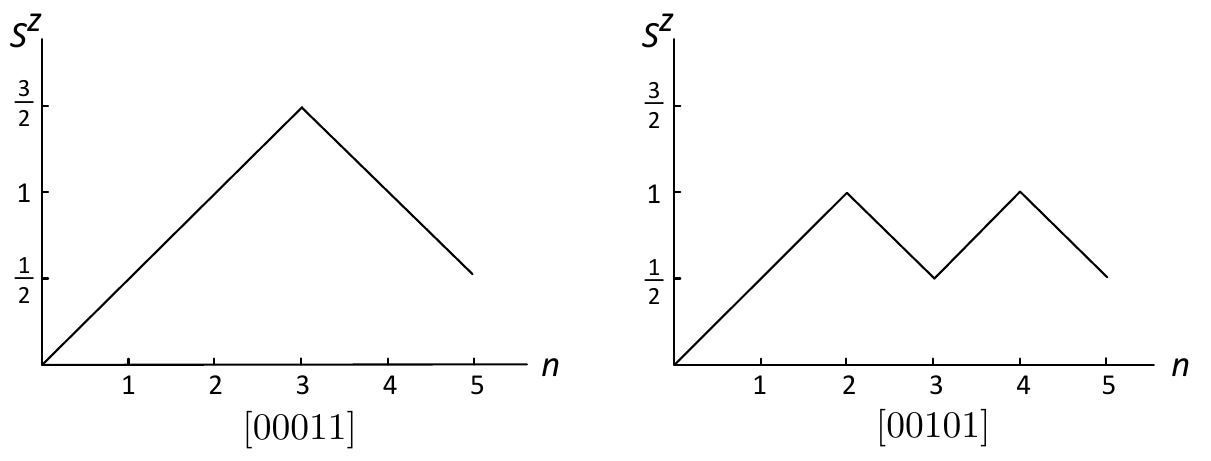}
\caption{Two path diagrams and the corresponding primitive spin functions $p_n$ for $n=5,S^z=1/2$. Taken from~\cite{Pauncz-2000}.}\label{fig.path-diagram}
\end{figure}

The abstract primitive spin function $\vec p_n$ is then intuitively presented by the corresponding primitive path.
Now we would ask the natural question: how to generate all the legal primitive paths terminating at $(n,S^z_n)$~
(Exercise 2.5 of~\cite{Pauncz-2000})?

Again we could do this forwardly or backwardly, and the latter would be more efficient. The backward approach could be implemented
recursively based on the following rules:
\begin{framed}
\noindent
\emph{\hspace*{1em}  1. Global rule: $|S^z_i|\le \frac{i}{2}$;\\
\hspace*{1em}  2. Local (branching) rule: $S^z_{i-1}=S^z_{i}\pm 1/2$.}
\end{framed}
Once all the spin projections $S^z_i$ are specified, the path diagram symbol $\vec \rho_n$ could be obtained through (\ref{eq:edge-spin3}).

The above problem could be of formal value. The more practical question would be, given the spin states $|S_n,S_n^z,\vec \lambda_n\rangle$,
how to generate all the legal primitive spin functions $\vec p_n$, or path diagram symbols $\vec \rho_n$? This can be considered as a constrained version of the above
problem. And the solution could be obtained by adapting the corresponding procedure. Again we use the backward approach, and generate the spin
projections recursively according to:
\begin{framed}
\noindent
\emph{\hspace*{1em}  1. Projection rule: $|S^z_i|\le S_i$;\\
\hspace*{1em}  2. Local (branching) rule: $S^z_{i-1}=S^z_{i}\pm 1/2$.}
\end{framed}
Notice that the projection rule would guarantee that the global rule $|S^z_i|\le \frac{i}{2}$ be satisfied. Again from all the spin
projections $S_i^z$ we can derive the path diagram symbols $\vec \rho_n$, and then the primitive spin functions $\vec p_n$.

In this way, we could enumerate all the components $\vec p_n$ of the given spin states $|S_n,S_n^z,\vec \lambda_n\rangle$. To completely
determine the state, we need to calculate the corresponding amplitudes of them. Such an algorithm has been developed long ago~\cite{GTG-1976},
and succinctly described in~\cite{Pauncz-2000}. Below we rewrite the algorithm in our formalism, so that the relation to our SCS-CG circuits
would be transparent.

Recall that at each step of the branching path, we have the decompositions (\ref{eq:spin-dec1}) and (\ref{eq:spin-dec2}),
depending on the value of $\vec \lambda_i[i]$. Namely, when $\vec \lambda_{i}[i]=+1$, we have
\begin{equation}
|S_{i},S_{i}^z,\vec \lambda_{i}\rangle=\sqrt{\frac{S_i+S_i^z}{2S_i}}|S_i-\frac{1}{2},S_{i}^z-\frac{1}{2},\vec \lambda_{i-1}\rangle\otimes|0\rangle+
\sqrt{\frac{S_i-S_i^z}{2S_i}}|S_i-\frac{1}{2},S_{i}^z+\frac{1}{2},\vec \lambda_{i-1}\rangle\otimes|1\rangle.\label{eq:spin-dec3}
\end{equation}
And when $\vec \lambda_{i}[i]=-1$, we have
\begin{equation}
|S_{i},S_{i}^z,\vec \lambda_{i}\rangle=-\sqrt{\frac{S_i-S_i^z+1}{2S_i+2}}|S_i+\frac{1}{2},S_{i}^z-\frac{1}{2},\vec \lambda_{i-1}\rangle\otimes|0\rangle+
\sqrt{\frac{S_i+S_i^z+1}{2S_i+2}}|S_i+\frac{1}{2},S_{i}^z+\frac{1}{2},\vec \lambda_{i-1}\rangle\otimes|1\rangle.\label{eq:spin-dec4}
\end{equation}
Within each decomposition, the two terms obviously correspond to $\vec p_n[i]=0$ and  $\vec p_n[i]=1$ respectively. We could therefore
write the four coefficients in a succinct way as:
\begin{eqnarray}
C(0,0;S,S^z)&=&[(S+S^z)/(2S)]^{1/2}\nonumber\\
C(0,1;S,S^z)&=&[(S-S^z)/(2S)]^{1/2}\nonumber\\
C(1,0;S,S^z)&=&-[(S-S^z+1)/(2S+2)]^{1/2}\nonumber\\
C(1,1;S,S^z)&=&[(S+S^z+1)/(2S+2)]^{1/2}.\nonumber
\end{eqnarray}
Here the first parameter of the function is the current element of $\vec b_n$, and and the second parameter is the current
element of $\vec p_n$. When both $\vec b_n$
and $\vec p_n$ are fixed, one of the four coefficients would be selected. The total amplitude are then given by the product of all the
selected coefficients along the branching path:
\begin{equation}
A(\vec b_n;\vec p_n)=\prod_{i=2}^n~C(\vec b_n[i],\vec p_n[i];S_i,S_i^z).\label{eq:BP-number}
\end{equation}
%Here all the spins $S_i$ could be obtained from the branching diagram symbols $\vec \lambda_n$ or $\vec b_n$, and all the spin projections
%$S_j^z$ could be obtained from the path diagram symbols $\vec \rho_n$ or $\vec p_n$.
Enumerating all the path diagram symbols $\vec p_n$
and the corresponding amplitudes $A(\vec b_n;\vec p_n)$, we can reconstruct $|S_n,S_n^z,\vec \lambda_n\rangle$.

We would like to make some remarks on the above reconstruction procedure. First of all, the whole algorithm, especially the final amplitude
(\ref{eq:BP-number}), could be considered as the classical counterpart of the SCS-CG circuits (\ref{eq:CG+},\ref{eq:CG-}) and their recursive composition (\ref{eq:U-path2}).
In some sense, the final amplitude (\ref{eq:BP-number}) could just be obtained from (\ref{eq:U-path2}) by making a superficial
quantum measurement. After the measurement, the whole ``path integral'' decoheres, and collapses to individual classical paths. Secondly, the number of path diagram symbols for a given
branching path $\vec \lambda_n$ could be very large, depending on $\vec \lambda_n$ and $S_n^z$. Therefore, in general such an algorithm
can not be efficient. However, we could still use it to verify our quantum circuits for small number of qubits.
Thirdly, the coefficients appearing in the above two decompositions could vanish apparently. However, by carefully check the situation
with vanishing coefficients, one could see that the corresponding primitive states always violate the projection rule. In other words,
these otherwise nonphysical states are exactly killed with vanishing coefficients. Therefore,
if the path diagram symbols $\vec p_n$ is generated with the projection rules and local branching rules satisfied, the final amplitude
$A(\vec b_n;\vec p_n)$ would always be nonzero. This would be one of the many benefits of the backward generation procedure. The
allowed primitive spin functions are discussed in detail in~\cite{Pauncz-2000}.

\subsubsection{Classical Evaluation on Spin Trees}

Since the branching paths could be directly lifted into the branching spin trees, we would like to generalize the above construction
to arbitrary spin trees. It turns out that, this could done in a very natural way.

We start from such a question: given the underlying tree graph $T^n$ and the output spin $S$, how to enumerate all the legal spin
assignments $\mathcal S$? Now we know it would be better to do this in a backward way. Here the depth of the tree could be taken as
(reversed) time. So starting from the node incident with the root edge, we perform spin decomposition on each node recursively
 to the rules:
\begin{framed}
\noindent \emph{\hspace*{1em}  1. Global rule: the root spin for the subtree $ST^m$ takes value from
$\{0,1,2,...,m/2\}$ when $m$ is even, and from $\{1/2,3/2,...,m/2\}$ when $m$ is odd;\\
\hspace*{1em}  2. Local (branching) rule: the three spins on the two incoming edges and the outgoing edge satisfy the triangular rule
in (\ref{eq:Triangle-Integer}).}
\end{framed}
Notice that we do not include the integer rule here, because it could be implied immediately by the other rules.

Now we turn to the classical counterpart. As before, we want to enumerate all the components of a given spin state. So we may ask the question:
giving a specific spin tree $ST^n$ and the corresponding state $|S,S^z,ST^n\rangle$, how to enumerate all its primitive spin functions?
We could do this in almost the same way as for branching diagram. To make the construction manifest, we first introduce
a refined notion of spin tree, a projected spin tree or a fibered spin tree:
\begin{framed}
\noindent
\textbf{Definition: Projected Spin Tree $pST^n:=\{ST^n,{\mathcal M}\}=\{n,T^n,({\mathcal S},{\mathcal M})\}$}:\\
\emph{\hspace*{1em} i)   $ST^n$ is a valid spin tree;\\
\hspace*{1em} ii)  ${\mathcal M}: E(T^n) \to \mathbb{Z}/2$ is a legal assignment of spin projection,
such that
\begin{equation}
{\mathcal M}(e)=S_e^z\in\{-S_e,-S_e+1,...,S_e\};
\end{equation}
\hspace*{1em} iii) On each node, the sum of incoming spin projections equals that of the outgoing one. That is,
the spin projections are conserved.}
\end{framed}
So for the projected spin tree, we have a spin pair $(S_e,S_e^z)$ on each edge. Due to the conservation law, we actually only need
the spin projections on the leaf edges to determine the whole assignment ${\mathcal M}$. Therefore a projected spin tree is just
a spin tree together with a legal path diagram symbol. Now our question could be enhanced to this:
giving a spin tree $ST^n$ and the corresponding state $|S,S^z,ST^n\rangle$, how to enumerate all the valid projected spin trees $pST^n$?

Essentially, we could accomplish this by making a proper measurement on the whole spin tree. Again we take the depth of the tree as
time. Starting from the node incident with the root edge, we recursively measure each node and make the assignment of spin projections
according to the following rules:
\begin{framed}
\noindent\emph{\hspace*{1em} 1. Projection rule: the spin projection $S_e^z$ on each incoming edge $e$ takes value from $\{-S_e,-S_e+1,...,S_e\}$;\\
\hspace*{1em} 2. Local (branching) rule: the spin projections are conserved.}
\end{framed}
Once we obtain all the valid projected spin trees, we immediately get all the primitive spin functions. More precisely, the spin projections
on the leaf edges are simply $\vec \rho_n[i]/2$, which can be converted into $\vec p_n$.

The last question would then be, for a legal projected spin tree $pST^n$, how to calculate its amplitude in the objective state
$|S,S^z,ST^n\rangle$? It turns out the result could be directly read out from the projected spin tree $pST^n$. For each node $n_j$ in $pST^n$,
we now have two incoming spin pairs, say $(S_{jL},S^z_{jL})$ and $(S_{jR},S^z_{jR})$, and one outgoing spin pair $(S_j,S_j^z)$. So we immediately
get a unique CG coefficient:
\begin{equation}
\mathrm{ev}[\mathrm{CG}](n_j)=\langle S_{jL},S_{jL}^z;S_{jR},S_{jR}^z|S_j,S_j^z\rangle.
\end{equation}
The full amplitude is simply the product of all these CG coefficients:
\begin{equation}
A(ST^n;pST^n)=\prod_{j=1}^{n-1}~ \mathrm{ev}[\mathrm{CG}](n_j).\label{eq:A-Tree}
\end{equation}
In the case of a branching tree, this formula reproduces the amplitude (\ref{eq:BP-number}). In fact, it not only greatly generalizes
the branching path amplitude (\ref{eq:BP-number}), but also provides a very nice geometric picture. In the case of branching diagram,
branching paths and primitive paths are two separate geometric objects, defined on different plots. Here the path diagram symbol is
naturally integrated in to the spin tree. However, a nice property of
the amplitude (\ref{eq:BP-number}) is lost in such a generalization. The amplitude (\ref{eq:A-Tree}) is no longer guaranteed to be nonzero
for every legal projection. For example, the following coupling
\begin{equation}
 (1,0) \otimes (1,0) \to (1,0)
\end{equation}
is clearly legal, but has a vanishing CG coefficient by accident.

In some sense, the above amplitude could be considered as collapsed from the full quantum operator (\ref{eq:U-CG}). Since the number of valid projections could increase very quickly, such a classical
construction could not be always efficient. Still, we could use it to verify the quantum circuits for small qubit numbers. With the above two classical algorithms, we have made a huge number of verifications of the spin states prepared with our SCS-CG circuits and
WDB-CG circuits. Of course, our quantum circuits withstand all these tests. We omit all these tedious processes and results.

\section{Summary and Outlook}

As Feynman put it, quantum processes could be difficult to simulate on classical computers, and it would be natural to simulate them on quantum
computers. , As an ubiquitous quantum process in nature, spin coupling should therefore be efficiently producible on quantum computers.
With this expectation in mind, we seek to prepare arbitrary spin eigenstates with quantum circuits, extending previous works of  B\"{a}rtschi
and Eidenbenz on Dicke states. The generalization turns out to be extremely successful. We successfully extend their SCS circuits
to prepare the canonical set of spin states produced from branching paths. We also successfully extend their WDB circuits to prepare all
the spin states produced from spin trees. The later generalization is rather nontrivial, and forces us to to extend the original WDB circuit
to the full weight range. Such a generalized WDB circuit could be of individual value. For example, they could be used to prepare a large class
of tree states defined in~\cite{Vittal-2025}.

In addition to these quantum circuits, we also develop their classical counterparts. That is, we give the classical algorithms for
evaluating the amplitudes of all the component states in a given spin state. Such an algorithm is proposed for the canonical
spin states from branching paths half a century ago
in the chemistry literature~\cite{GTG-1976}. We generalize it from branching paths to arbitrary spin trees, and endow it with a nice geometric picture.
We have used these classical algorithms to verify our circuits for small numbers of qubits. We believe these classical algorithms would also
 be helpful in practical calculations in quantum chemistry and quantum manybody physics.

There are many interesting directions to explore. Below we list a few of them.

First of all, as we mentioned, the original restricted WDB unitary $\mathrm{WDB}^k_{m,n-m}$ could be efficiently implemented
in depth ${\mathcal O}(k)$. In contrast, our implementation of the full WDB unitary $\mathrm{WDB}_{m,n-m}$ has a depth at least
${\mathcal O}(n^2)$. Also we have used an auxiliary qubit. How to simplify the circuit to at least linear depth? And how to get rid of
the auxiliary qubit? Since spin coupling is an intrinsic quantum process, we should be able to simulate it in an intrinsic way. That is to say,
in principle we should be able to eliminate the auxiliary qubit. However, such an intrinsic implementation would not necessarily lead to a
reduction of the circuit depth.

Secondly, our preparation circuits could serve as primitives for devising practical algorithms. The recently proposed DQI algorithm~\cite{DQI-2024} takes Dicke states as initial states. What kinds of algorithms could we develop with the other
kinds of spin states? Perhaps we could start with intrinsic quantum problems first. We have shown that the groundstate of the Heisenberg model
on complete bipartite graphs are Lieb-Mattis states.
%On bipartite graphs, Heisenberg model, or the equivalent QMC
%problem, could be transformed into the so-called Einstein-Podolsky-Rosen~(EPR) problem~\cite{King-2023}.
%EPR problem is also referred to as ferromagnetic XXZ model in the physics literature.
In fact, for complete graphs and complete bipartite graphs, both QMC and the related Einstein-Podolsky-Rosen~(EPR) problem~\cite{King-2023}
could be completely solved~\cite{CM-2013,APS-2025}. The eigenstates are all given by spin states. One could continue this hierarchy of
graphs by considering graphs represented by alternating sums of cliques, as recently proposed in~\cite{Watts-2023}. Such a clique decomposition~\cite{Watts-2023}
naturally generates a rooted tree, though not necessarily a binary one. Could we adjust our WDB-CG circuits to solve QMC and/or EPR on this
hierarchy of graphs, at least approximately? This sounds like a very promising direction to pursue.

Finally, instead of preparing individual spin states, a more ambitious goal would be to implement the quantum Schur transformation.
As we know, branching diagram provides a canonical basis of spin eigenstates, which are orthogonal and complete. Could we collect all the spin
eigenstates produced from branching paths, and achieve the whole Schur transformation? We believe that this could be done in principle.
In fact, the quantum algorithm proposed in ~\cite{Bacon-2006,Bacon-2007} for the Schur transformation implicitly used branching paths.
Could we integrate the concrete circuits in this paper into the procedure in~\cite{Bacon-2006,Bacon-2007}, and get a more transparent and
practical realization of the Schur transformation? Such an implementation could greatly broaden the use of Schur transformation in algorithm design, as expected in \cite{Bacon-2007}.

%\newpage

\appendix

\section{Angles in WDB-CG circuits}\label{App.angle}

In this appendix we give the derivation of all the angles in the WDB-CG circuits, so that they would be complete.
We deal with the WDB module and
the negative-logic WDB module separately.

\subsection{Angles in $\mathrm{WDB}$}
In the WDB block, we would like to realize the following duplication transformation for various $l$:

\begin{equation}
|\psi_0\rangle = |0^l\rangle |1^l\rangle \to |\psi_1\rangle=\sum_{i=0}^l c_i |0^{l-i}1^i\rangle|1^l\rangle. \label{eq.duplication-1}
\end{equation}

We achieve it with the following circuit:

\begin{figure}[h]
\centering
	\includegraphics[width=\textwidth]{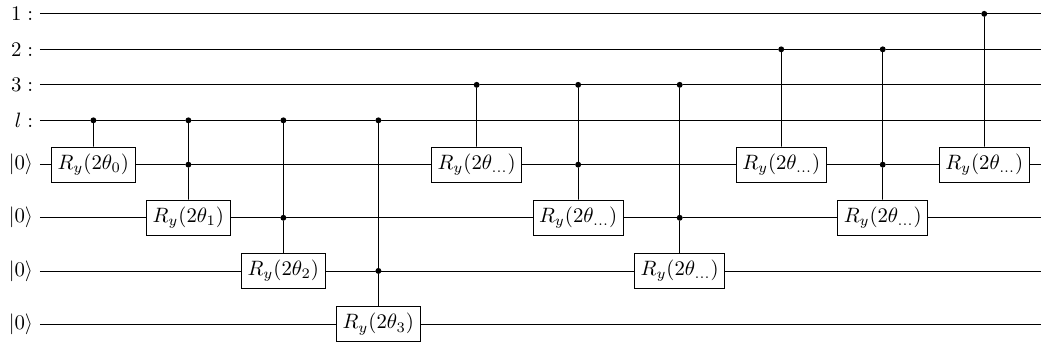}
\caption{Duplication module of the WDB circuit. Taken from~\cite{BE-2022A}.}\label{fig.dup2}
\end{figure}

Since the circuit is modularized, we may focus on a fixed $l$. The duplication part for a fixed $l$ consists of
a series of $l$ controlled $R_y(2\theta)$ rotations. Let us denote the $l$ angles as $\theta_i, i=0,...,l-1$, with $i$ increasing with time order.
We also take $\theta_l=0$. Now we expand all the rotation gates, and get:
\begin{equation}
|\psi_1\rangle=\cos \theta_0|0^l\rangle|1^l\rangle+\sin \theta_0 \cos \theta_1 |0^{l-1}1\rangle|1^l\rangle+\sin \theta_0\sin \theta_1\cos\theta_2|0^{l-2}1^2\rangle|1^l\rangle+...+\prod_{j=0}^{l-1}\sin \theta_j|1^l\rangle|1^l\rangle.
\end{equation}
Equating it to our objective state (\ref{eq.duplication-1}), we obtain:
\begin{eqnarray}
c_0&=&\cos \theta_0,\nonumber\\
c_i&=&\cos \theta_i\prod_{j=0}^{i-1}\sin \theta_j,\quad \forall~ 1\le i\le l.
\end{eqnarray}
Notice that we have used $\theta_l=0$ to keep the expression universal. Taking the quotient of successive coefficients, we get
\begin{eqnarray}
c_0&=&\cos \theta_0,\nonumber\\
\frac{c_{i+1}}{c_i}&=&\frac{\sin \theta_i}{\cos \theta_i}\cos \theta_{i+1},\quad \forall~ 0\le i\le l-1.
\end{eqnarray}
Now we take all the sines to be positive. Then the signs of all $\cos \theta_i$'s could be determined directly
from $c_i$. The values of them could be determined recursively. For example,
$\cos \theta_0$ is given by $c_0$, $\cos \theta_1$ is then fixed by $\frac{c_{1}}{c_0}$ and $\frac{\sin\theta_0}{\cos \theta_0}$ together,
and so on.

In some cases, some coefficients may vanish by accident. Then we could not naively take the quotient. This is easy to handle. Say $c_i=0$, then
we could just set $\theta_i=\pi/2$. After singling out these vanishing coefficients, we then solve the remaining angles as
above.

\subsection{Angles in $\overline{\mathrm{WDB}}$}

We also need to fix the rotation angles in the negative-logic WDB module. A sketch of such a module is shown below in Fig.~\ref{fig:anti-WDB}.

\begin{figure}[h]
\centering
	\includegraphics[width=\textwidth]{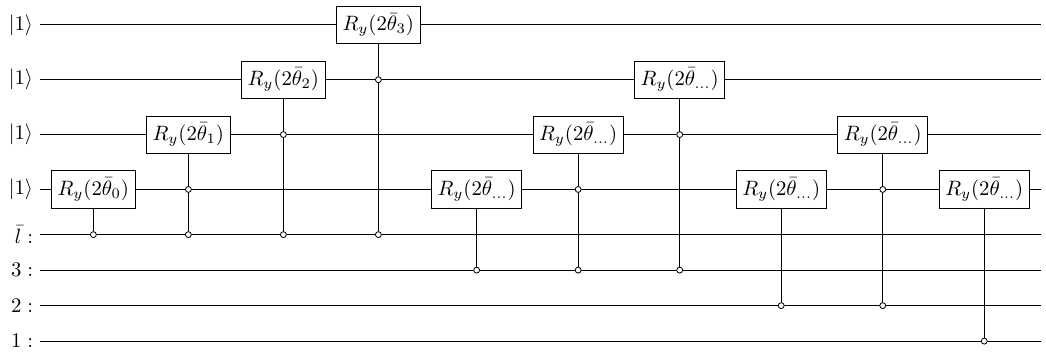}
\caption{Duplication module of the negative-logic WDB circuit.}\label{fig:anti-WDB}
\end{figure}

Now we want to perform the following transformation:
\begin{equation}
|{\bar \psi}_0\rangle = |0^{\bar l}\rangle |1^{\bar l}\rangle \to |{\bar \psi}_1\rangle=\sum_{ i=0}^{\bar l} {\bar c}_i |0^{\bar l}\rangle|0^i1^{\bar l-i}\rangle. \label{eq.duplication-2}
\end{equation}
We achieve this with a series of $\bar l$ controlled $R_y(2 {\bar \theta})$ rotations. The angles are denoted as ${\bar \theta}_i, i=0,...,{\bar l}-1$ respectively.
We also set ${\bar \theta}_{\bar l}=0$. Now applying all these gates on  $|{\bar \psi}_0\rangle$, we get
\begin{eqnarray}
|\bar{\psi}_1\rangle&=&\cos {\bar \theta}_0 |0^{\bar l}\rangle |1^{\bar l}\rangle-\sin {\bar \theta}_0\cos {\bar \theta}_1|0^{\bar l}\rangle |0^11^{\bar l-1}\rangle+\sin {\bar \theta}_0\sin {\bar \theta}_1\cos {\bar \theta}_2|0^{\bar l}\rangle |0^21^{\bar l-2}\rangle \nonumber\\
&&+...+(-1)^{\bar l}\prod_{j=0}^{\bar l-1}\sin {\bar \theta}_j |0^{\bar l}\rangle |0^{\bar l}\rangle.
\end{eqnarray}
Comparing it with (\ref{eq.duplication-2}) term by term, we obtain
\begin{eqnarray}
\bar c_0&=&\cos {\bar \theta}_0,\nonumber\\
\bar c_i&=&(-1)^i\cos {\bar \theta}_i\prod_{j=0}^{i-1}\sin {\bar\theta}_j,\quad 1\le i\le \bar l,
\end{eqnarray}
where we have used ${\bar \theta}_{\bar l}=0$. Again we take the quotient of successive coefficients, and get
 \begin{eqnarray}
\bar c_0&=&\cos \bar \theta_0,\nonumber\\
\frac{\bar c_{i+1}}{\bar c_i}&=&-\frac{\sin \bar \theta_i}{\cos \bar \theta_i}\cos \bar \theta_{i+1},\quad 0\le i\le \bar l-1,
\end{eqnarray}
Now we set all sines to be negative, then the sign of each cosine $\cos \bar \theta_i$
is determined by that of $\bar c_i$. The value of each cosine is then fixed recursively. When some coefficients vanish by accident,
we single them out, and set the corresponding angles to be $\pi/2$. After that we solve the remaining angles recursively as described above.

%\end{CJK*}


\begin{thebibliography}{10}

\bibitem{DQI-2024}
Stephen~P Jordan, Noah Shutty, Mary Wootters, Adam Zalcman, Alexander
  Schmidhuber, Robbie King, Sergei~V Isakov, Tanuj Khattar, and Ryan Babbush.
\newblock Optimization by decoded quantum interferometry.
\newblock {\em Nature}, 646(8086):831--836, 2025.

\bibitem{BE-2022}
Andreas B{\"a}rtschi and Stephan Eidenbenz.
\newblock Short-depth circuits for dicke state preparation.
\newblock {\em IEEE International Conference on Quantum Computing \&
  Engineering (QCE), 2022, pp. 87--96}, 2022.

\bibitem{BE-2019}
Andreas B{\"a}rtschi and Stephan Eidenbenz.
\newblock Deterministic preparation of dicke states.
\newblock In {\em International Symposium on Fundamentals of Computation
  Theory}, pages 126--139. Springer, 2019.

\bibitem{Carbone2022}
Alessandro Carbone, Davide~Emilio Galli, Mario Motta, and Barbara Jones.
\newblock Quantum circuits for the preparation of spin eigenfunctions on
  quantum computers.
\newblock {\em Symmetry}, 14(3):624, 2022.

\bibitem{Marti-2025}
Daniel Marti-Dafcik, Hugh~GA Burton, and David~P Tew.
\newblock Spin coupling is all you need: Encoding strong electron correlation
  in molecules on quantum computers.
\newblock {\em Physical Review Research}, 7(1):013191, 2025.

\bibitem{Vittal-2025}
Sunil Vittal, Anthony Wilkie, Nika Rastegari, Mostafa Atallah, and Rebekah
  Herrman.
\newblock Efficient circuits for leaf-separable state preparation.
\newblock {\em arXiv:2511.11227}, 2025.

\bibitem{Bacon-2006}
Dave Bacon, Isaac~L Chuang, and Aram~W Harrow.
\newblock Efficient quantum circuits for schur and clebsch-gordan transforms.
\newblock {\em Physical review letters}, 97(17):170502, 2006.

\bibitem{Bacon-2007}
Dave Bacon, Isaac~L Chuang, and Aram~W Harrow.
\newblock The quantum schur transform: I. efficient qudit circuits.
\newblock {\em Proceedings of the eighteenth annual ACM-SIAM symposium on
  Discrete algorithms (SODA), pp. 1235-1244}, 2007.

\bibitem{Baez-1995}
John~C Baez and James Dolan.
\newblock Higher-dimensional algebra and topological quantum field theory.
\newblock {\em Journal of mathematical physics}, 36(11):6073--6105, 1995.

\bibitem{VVS-1935}
J.~H. Van~Vleck and Albert Sherman.
\newblock The quantum theory of valence.
\newblock {\em Rev. Mod. Phys.}, 7:167--228, Jul 1935.

\bibitem{Pauncz-1977}
Ruben Pauncz.
\newblock Branching diagram and serber-type spin functions. algorithms for
  their construction and special properties.
\newblock {\em International Journal of Quantum Chemistry}, 12(2):369--382,
  1977.

\bibitem{Sugisaki-2016}
Kenji Sugisaki, Satoru Yamamoto, Shigeaki Nakazawa, Kazuo Toyota, Kazunobu
  Sato, Daisuke Shiomi, and Takeji Takui.
\newblock Quantum chemistry on quantum computers: A polynomial-time quantum
  algorithm for constructing the wave functions of open-shell molecules.
\newblock {\em The Journal of Physical Chemistry A}, 120(32):6459--6466, 2016.

\bibitem{Sugisaki-2019}
Kenji Sugisaki, Satoru Yamamoto, Shigeaki Nakazawa, Kazuo Toyota, Kazunobu
  Sato, Daisuke Shiomi, and Takeji Takui.
\newblock Open shell electronic state calculations on quantum computers: A
  quantum circuit for the preparation of configuration state functions based on
  serber construction.
\newblock {\em Chemical Physics Letters}, 737:100002, 2019.
\newblock Articles initially published in Chemical Physics Letters: X 1-4,
  2019.

\bibitem{Pauncz-2000}
Ruben Pauncz.
\newblock {\em The construction of spin eigenfunctions. An exercise book}.
\newblock Kluwer Academic/Plenum Publishers: New York, 2000.

\bibitem{Zeng-2013}
Zeng Jinyan.
\newblock {\em Quantum Mechanics (in Chinese), Fifth edition, volume 1}.
\newblock Science Press, 2013.

\bibitem{Penrose-1971}
Roger Penrose.
\newblock Angular momentum: an approach to combinatorial space-time.
\newblock {\em in Quantum Theory and Beyond, ed. T. Bastin, Cambridge
  University Press, Cambridge}, pages 151--180, 1971.

\bibitem{JB-2010}
John Baez and Mike Stay.
\newblock Physics, topology, logic and computation: a rosetta stone.
\newblock In {\em New structures for physics}, pages 95--172. Springer, 2010.

\bibitem{LM-1962}
Elliott Lieb and Daniel Mattis.
\newblock Ordering energy levels of interacting spin systems.
\newblock {\em Journal of Mathematical Physics}, 3(4):749--751, 07 1962.

\bibitem{CM-2013}
Toby Cubitt and Ashley Montanaro.
\newblock Complexity classification of local hamiltonian problems.
\newblock {\em SIAM Journal on Computing}, 45(2):268--316, 2016.

\bibitem{AGM-2020}
Anurag Anshu, David Gosset, and Karen Morenz.
\newblock {Beyond Product State Approximations for a Quantum Analogue of Max
  Cut}.
\newblock {\em Leibniz Int. Proc. Inf.}, 158:7:1--7:15, 2020.

\bibitem{APS-2025}
Anuj Apte, Ojas Parekh, and James Sud.
\newblock {Conjectured Bounds for 2-Local Hamiltonians via Token Graphs}.
\newblock {\em arXiv:2506.03441}, June 2025.

\bibitem{Watts-2023}
Adam~Bene Watts, Anirban Chowdhury, Aidan Epperly, J.~William Helton, and Igor
  Klep.
\newblock {Relaxations and Exact Solutions to Quantum Max Cut via the Algebraic
  Structure of Swap Operators}.
\newblock {\em Quantum}, 8:1352, 2024.

\bibitem{BBKL-2026}
Ainesh Bakshi, Arpon Basu, Pravesh Kothari, and Anqi Li.
\newblock Sharp bounds on the eigenvalues of kikuchi graphs and applications to
  quantum max cut.
\newblock {\em arXiv:2605.14994}, 2026.

\bibitem{GTG-1976}
J.~E. Grabenstetter, T.~J. Tseng, and Friedrich Grein.
\newblock Generation of genealogical spin eigenfunctions.
\newblock {\em International Journal of Quantum Chemistry}, 10:143--149, 1976.

\bibitem{King-2023}
Robbie King.
\newblock An improved approximation algorithm for quantum max-cut on
  triangle-free graphs.
\newblock {\em Quantum}, 7:1180, 2023.

\end{thebibliography}

\begin{thebibliography}{100}

\bibitem{BE-2022A}
Andreas B{\"a}rtschi and Stephan Eidenbenz.
\newblock Short-depth circuits for dicke state preparation.
\newblock {\em IEEE International Conference on Quantum Computing \&
  Engineering (QCE), 2022, pp. 87--96}, 2022.



\end{thebibliography}
\end{document}